\documentclass[11pt, a4paper]{article}
\usepackage{jheparxiv}
\usepackage[utf8]{inputenc}
\usepackage{amsmath}
\usepackage{amsfonts}
\usepackage{amssymb}
\usepackage{latexsym}
\usepackage{mathrsfs}
\usepackage{braket}		%Bra Ket Notation
\usepackage{setspace}
\usepackage{graphicx}
\usepackage{color}
\usepackage{xcolor}
\usepackage{slashed}
\usepackage{twistor}
\usepackage[all]{xy}
\usepackage{tikz-cd}
\usepackage{mathtools}
\usepackage{tikz}

\usepackage{graphicx}

\newcommand{\cK}{\mathcal{K}}

\newcommand{\qed}{\hfill \ensuremath{\Box}}

\newcommand{\eps}{\epsilon}

\newcommand{\msf}[1]{\mathsf{#1}}

\title{Constraining higher-spin $S$-matrices}

\author[a]{Tung Tran}
\affiliation[a]{
 Service de Physique de l’Univers, Champs et Gravitation,\\
Université de Mons, 20 place du Parc, 7000 Mons, Belgium
}%

\emailAdd{vuongtung.tran@umons.ac.be}

\abstract{There are various no-go theorems that tightly constrain the existence of local higher-spin theories with non-trivial $S$-matrix in flat space. Due to the existence of higher-spin Yang-Mills theory with non-trivial scattering amplitudes, it makes sense to revisit Weinberg's soft theorem -- a direct consequence of the Lorentz invariance of the $S$-matrix that does not take advantage of unitarity and parity invariance. By working with the chiral representation -- a representation originated from twistor theory, we show that Weinberg's soft theorem can be evaded and non-trivial higher-spin $S$-matrix is possible. In particular, we show that Weinberg's soft theorem is more closely related to the number of derivatives in the interactions rather than spins. We also observe that all constraints imposed by gauge invariance of the $S$-matrix are accompanied by polynomials in the soft momentum of the emitted particle where the zeroth order in the soft momentum is a charge conservation law.}

\begin{document}

\maketitle
  
%%%%%%%%%%%%%%%%%%%%%%%%%%%%%%%%%%%%%%%%%%
\section{Introduction}\label{sec:1}
Recently, the first example of a higher-spin theory with non-trivial $S$-matrix has been found~\cite{Adamo:2022lah} despite the no-go theorems in flat spacetime \cite{Weinberg:1964ew,Coleman:1967ad} and (A)dS \cite{Maldacena:2011jn,Boulanger:2013zza,Sleight:2017pcz} -- see e.g. \cite{Weinberg:1995mt,Bekaert:2010hw,Schwartz:2014sze,Strominger:2017zoo,McLoughlin:2022ljp} for a review on no-go theorems, and \cite{Bekaert:2022poo,Ponomarev:2022vjb} for a review on higher-spin theories. As argued in \cite{Adamo:2022lah}, higher-spin Yang-Mills (HS-YM) can evade no-go theorems since it is intrinsically chiral and not a parity-invariant theory. This is in agreement with the common knowledge in higher-spin gravities (HSGRAs): constructing toy models of HSGRAs requires letting go of at least one of the important features of field theory such as unitarity or locality. Essentially, unitary HSGRAs are known to be non-local \cite{deMelloKoch:2010wdf,Bekaert:2015tva,Boulanger:2015ova,Aharony:2022feg}, and local higher-spin theories are known to be non-unitary; and there seems to be no compromise.

While giving up parity-invariance is reasonable,\footnote{Some simple theories such as self-dual Yang-Mills \cite{Bern:1993qk,Mahlon:1993si,Bardeen:1995gk,Bern:1996ja} and self-dual gravity \cite{Bern:1998sv,Krasnov:2016emc} have broken unitary and parity. Nevertheless, they are consistent truncation of full Yang-Mills and gravity theories.} abandoning locality often carries the risk of having pathological theories. In particular, perturbative methods such as the light-front approach \cite{Bengtsson:1983pd,Bengtsson:1986kh,Metsaev:2005ar} or Noether procedure (cf., \cite{Barnich:1993vg,Manvelyan:2010jr,Joung:2013nma}) will be ill-defined in this case.\footnote{If non-locality is allowed, there will be no obstruction to constructing higher-order gauge-invariant vertices that preserve gauge symmetries.} For this reason, if we want to keep locality at all costs, all viable higher-spin theories we can have are either (quasi-)topological \cite{Blencowe:1988gj,Bergshoeff:1989ns,Pope:1989vj,Fradkin:1989xt,Metsaev:1991mt,Metsaev:1991nb,Campoleoni:2010zq,Henneaux:2010xg,Gaberdiel:2010pz,Gaberdiel:2012uj,Gaberdiel:2014cha,Grigoriev:2019xmp,Grigoriev:2020lzu,Ponomarev:2016lrm,Metsaev:2018xip,Metsaev:2019dqt,Metsaev:2019aig,Krasnov:2021nsq,Tran:2022tft,Tsulaia:2022csz,Herfray:2022prf}, higher-spin extension of conformal gravity~\cite{Segal:2002gd,Tseytlin:2002gz,Bekaert:2010ky}, or quasi-chiral \cite{Sperling:2017dts,Sperling:2018xrm,Steinacker:2022jjv,Adamo:2022lah}.\footnote{Here, by quasi-chiral, we mean theories that are chiral in nature but can have non-trivial $S$ matrices. They are expected to be consistent truncations of some unitary theories whose descriptions require relaxing locality.} Note that all known (quasi-)chiral higher-spin theories have complex action functionals in spacetimes with Lorentzian signature.

As shown in \cite{Adamo:2022lah}, higher-spin Yang-Mills (HS-YM) theory has non-trivial scattering amplitudes in Lorentzian flat spacetime. This phenomenon can occur due to the fact that quasi-chiral theories are non-unitary and non-parity-invariant. Thus, they violate all assumptions of no-go theorems, which constrain higher-spin $S$-matrices. Hence, non-trivial scattering amplitudes are allowed. Very roughly, we can view this as a process of having many massless higher-spin particles scatter off of a higher-spin chiral `background'. To wit, we can view the chiral background as a deformation away from the Minkowski background \cite{Mason:2009afn,Adamo:2020yzi,Adamo:2022mev}; and consider a scattering process on this `non-trivial' background rather than a flat one. Lastly, it is worth noting that when studying chiral HSGRAs in (A)dS, even though the theory is chiral, it does not possess a trivial holographic $S$-matrix \cite{Skvortsov:2018uru}.\footnote{The 3-pt functions of chiral HSGRA in (A)dS do not match 3-pt functions of free CFTs.} For this reason, chiral HSGRA in (A)dS is expected to be dual to Chern-Simons matter theories \cite{Sharapov:2022awp}. See also recent development in Chern-Simons matter theories \cite{Jain:2022ajd} on the CFT side.

The non-triviality of quasi-chiral higher-spin theories is a compelling reason to revisit Weinberg's soft theorem \cite{Weinberg:1964ew} to investigate if there are other possible loopholes that can lead to the existence of these peculiar theories. In this work, we show that Weinberg's arguments can easily be evaded when working with the chiral representation \cite{Krasnov:2021nsq,Adamo:2022lah} -- a representation originating from twistor theory \cite{Penrose:1967wn}.\footnote{Different field representations feature different numbers of derivatives in the vertices. For example, it is well-known that the Fronsdal representation has more derivatives compared to the chiral one. See discussion in Section \ref{sec:2}.} 

For reference, let us recall the results of Weinberg, which are the following constraints:
\begin{align}
    \sum_i\mathtt{c}_{s,i}\,p_i^{\mu_1}\ldots p_i^{\mu_{s-1}}=0\,,
\end{align}
imposed by Lorentz invariance of the $S$-matrix. Here, $p_i^{\mu}$ is the momentum of the external leg $i$th and $s$ is the spin of the soft emitting particle. A direct consequence of the above is that the spin of the emitting soft particle cannot exceed two if we want to have non-trivial scattering amplitudes. This has partly extinguished the hopes of finding massless higher-spin theories in the past. 

The above problem can be overcome by working with the chiral representation. In this case, we observe that all constraints imposed by gauge invariance of the $S$-matrix in \cite{Weinberg:1964ew} will be accompanied by polynomials in soft momentum $k^{\alpha\dot\alpha}$, i.e. 
\begin{align}\label{mainresult}
   \parbox{90pt}{\includegraphics[scale=0.2]{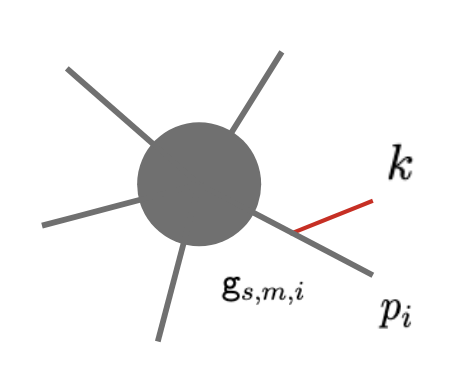}}\Leftrightarrow\qquad  0=\sum_m\big(\sum_i^n\tg_{s,m,i}\,\underbrace{p_{i\,\alpha\dot\gamma}\ldots p_{i\,\alpha\dot\gamma}}_{m\ 
   \text{times}}\big)\underbrace{k_{\alpha}{}^{\dot\gamma}\ldots k_{\alpha}{}^{\dot\gamma}}_{m\ \text{times}}\,.
\end{align}
Here, $m$ is the number of derivatives in each cubic vertex and $p_i^{\alpha\dot\alpha}$ is the momentum of the external leg $i$th. At $m=0$, we find charge conservation, while at $m=1$, we recover the equivalence principle where $\tg_{s,1,i}=const$ to avoid trivial scattering. In contrast with the result of \cite{Weinberg:1964ew}, we do not need to trivialize the coupling constants $\tg_{s,m,i}$ to zero at higher-order in derivatives when the soft limit $k^{\alpha\dot\alpha}\rightarrow 0$ is strictly applied. This leads us to an intriguing conclusion that the decisive factor for non-trivial higher-spin scatterings is not spin but the number of derivatives in the interactions. To support this observation, we also compute higher-spin soft factors to obtain further constraints for macroscopic higher-spin fields. It is important to emphasize that by choosing to work with the chiral representation, we will have a discrimination between the positive and negative helicity fields. As a consequence, we find that all conservation laws come from positive helicity soft particles. 

\medskip

This note is structured as follows. Section \ref{sec:2} provides some useful information on field representations, and Weinberg's soft theorem. Next, we study the infrared (IR) physics of soft emitting higher-spin particles in Section \ref{sec:3}. Various higher-spin soft factors and their implications are presented in the same section. Finally, we conclude in Section \ref{sec:4}.

\paragraph{Notation.} Throughout this note, we adapt the same convention as in \cite{Adamo:2022lah}.

%%%%%%%%%%%%%%%%%%%%%%%%%%%%%%%%%%%%%%%%%% 
%%%%%%%%%%%%%%%%%%%%%%%%%%%%%%%%%%%%%%%%%%

\section{Provision}\label{sec:2}

\subsection{Fronsdal representation vs chiral representation}\label{Fronsdalvschiral}
Let us set the stage for our discussion regarding higher-spin soft interactions in flat space by reviewing the field representations used in this paper.

\begin{figure}[ht]
    \centering
    \includegraphics[scale=0.33]{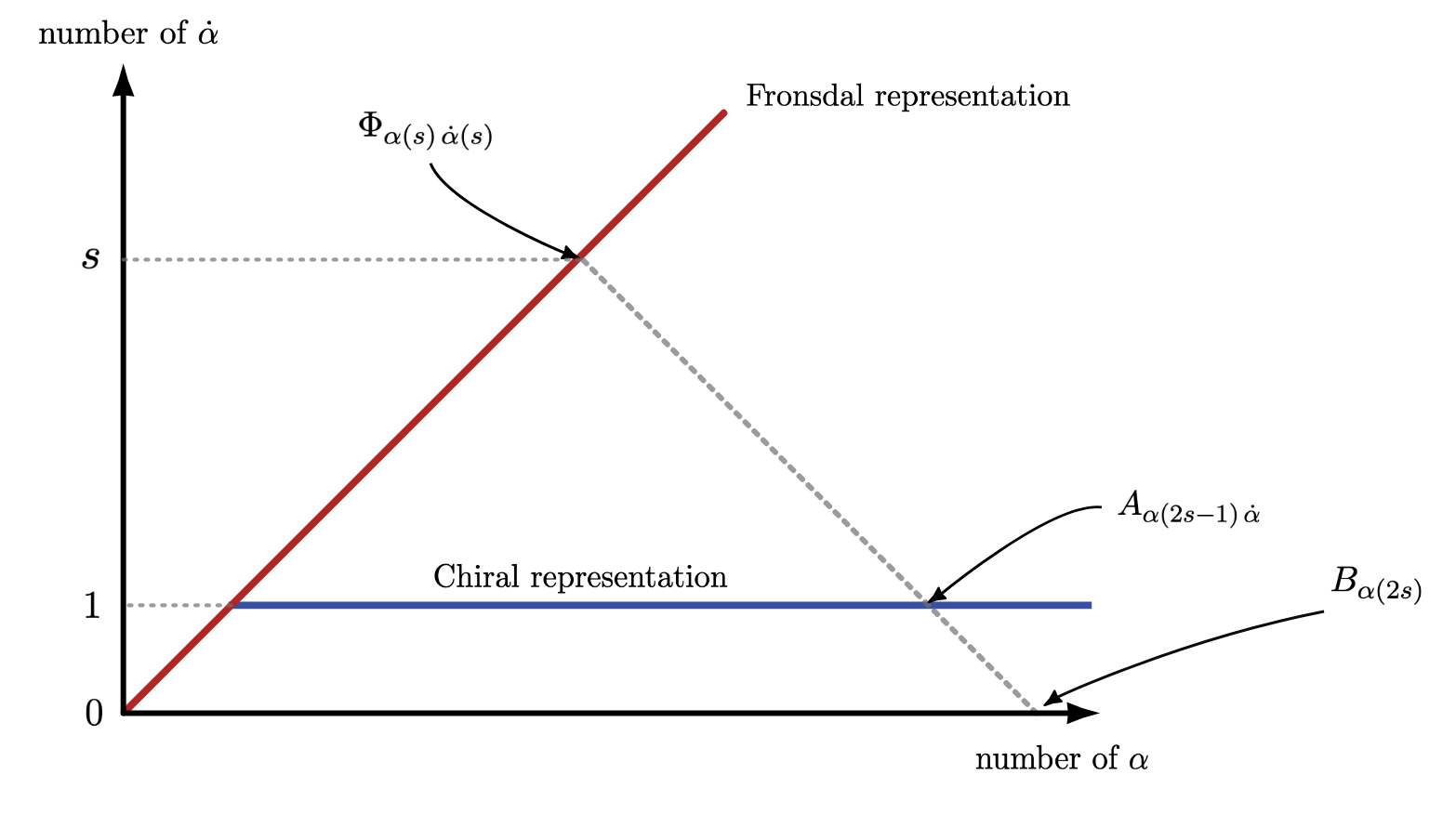}
    \caption{A spin-$s$ gauge field in two different representations: Fronsdal (red) vs chiral (blue) \cite{Adamo:2022lah}.}
\end{figure}

As is well-known, a totally symmetric rank-$s$ tensorial field in $4d$ can be written as a rank-$2s$ spin-tensor $T_{\alpha(s)\,\dot\alpha(s)}$ which is an element of spinor representation space $\mathtt{S}(s,s)$. Here, the first argument stands for the number of un-dotted/negative chirality $SL(2,\C)$ spinor indices and the second argument marks the number of dotted/positive chirality spinor indices. The representation $\mathtt{S}(s,s)$ is also known as the balanced/Fronsdal representation, where fields are Lorentzian-real. To date, this is the most studied representation in the higher-spin literature (see e.g. \cite{Sezgin:2017jgm,Skvortsov:2022wzo}). Although being Lorentzian-real is a desired feature for unitarity, theories constructed from the Fronsdal representation suffer from non-locality issues starting from quartic higher-spin interactions \cite{Bekaert:2015tva,Boulanger:2015ova,Skvortsov:2015lja}. For a more intuitive view, let us consider the cubic vertex in the Fronsdal representation~\cite{Fronsdal:1978rb,Boulanger:2015ova,Sharapov:2022awp,Sharapov:2022faa,Conde:2016izb}:
\small
\begin{align}\label{Fronsdalcubic}
    V_3=\sum_{m,s_i}C_m^{s_1,s_2,s_3}\partial^m\Phi_{s_1}\Phi_{s_2}\Phi_{s_3}\,,\qquad m=\sum_{i=1}^3s_i-2\text{min}(s_1,s_2,s_3)\,.
\end{align}
\normalsize
where we do not wish to specify how indices are contracted. Here, $\Phi_s\equiv\Phi_{\alpha(s)\,\dot\alpha(s)}$ are known as the Fronsdal spin-$s$ fields and $C_m^{s_i}$ is some coupling constant that scales with number of derivatives $m$ and spins $s_i$. Note that, we can recover only a sub-sector of cubic vertices for higher-spin fields in \cite{Bengtsson:1983pd,Metsaev:1991mt,Metsaev:1991nb} from \eqref{Fronsdalcubic}. 
Furthermore, it is well-known that one cannot construct local interactions starting from the quartic with the Fronsdal representation, see e.g. \cite{Roiban:2017iqg}.\footnote{See also recent attempts to tame non-locality of unitary higher-spin theories by defining new diagram rules for the holographic $S$-matrix \cite{Lysov:2022nsv,Neiman:2022enh}.}

\medskip

The insufficiency of the Fronsdal representation demands a different field realization to avoid non-locality issues, and to reproduce all cubic vertices available in the light-cone gauge. One observes that the \textit{chiral representation} used in the twistor construction for action functionals of local higher-spin theories \cite{Krasnov:2021nsq,Tran:2022tft} can satisfy the above requirement since the chiral representation is known to produce the lowest number of derivatives when constructing HSGRAs. For instance, in the chiral representation, the $(-,+,+)$ cubic vertex of massless higher-spin fields reads (schematically):
\begin{align}
   V_3^{-,+,+}= \sum_m\, C_{m,s}^{s_1,s_2}B_{\alpha(2s)}\,\underbrace{\p_{\alpha \dot\gamma}\ldots \p_{\alpha\dot\gamma}}_{m\ \text{times}}A^{\alpha(2s_1-1)\,\dot\alpha}\,\p_{\alpha}{}^{\dot\gamma}\ldots \p_{\alpha}{}^{\dot\gamma}A^{\alpha(2s_2-1)}{}_{\dot\alpha}\,,
\end{align}
where contraction between un-dotted indices forces $s+m=s_1+s_2-1$. In addition, the un-dotted indices in the derivatives are understood to be contracted with the un-dotted indices of physical fields $A$ in all possible ways. It can be checked that this type of contraction can reproduce all cubic vertices in the light-cone gauge following the procedure in \cite{Krasnov:2021nsq,Tran:2021ukl}. As an observation, we would like to emphasize that there are many field representations that can carry the same degrees of freedom. However, depending on how we use them, we will have different number of derivatives in the interactions.

\medskip

Of course, nothing comes for free. In the chiral representation space $\mathtt{S}(2s-1,1)$, higher-spin gauge fields are intrinsically chiral and not Lorentzian-real. As a consequence, theories constructed from this representation will, in general, break parity-invariance. Nevertheless they are consistent truncation of full unitary theories, see e.g. \cite{Krasnov:2016emc,Krasnov:2020lku} for the `chiral' pure connection formulation for General Relativity. As a result, this type of (quasi-)chiral theories can evade all no-go theorems while they enjoy having non-trivial $S$-matrices in flat space (see the first example in \cite{Adamo:2022lah}).\footnote{It is worth noting that, even though (quasi-)chiral theories have complex action functionals, their observables might still be unitary in the sense that they are part of a larger set of amplitudes which form a unitary $S$-matrix. For example, although SDYM and SDGRA have vanishing tree-level amplitudes, their non-trivial all-plus one-loop amplitudes are also the amplitudes in YM and GR. Therefore, it is reasonable to expect that the observables of (quasi-)chiral theories belong to a set of amplitudes of some yet unknown unitary non-local higher-spin theories.}

\medskip

Although the higher-spin multiplet of bosonic (quasi-)chiral models is theory-dependent, it usually contains a tower of higher-spin generalizations of the Yang-Mills gauge potential $\bigcup_{s=1}^{\infty}\left\{A_{\alpha(2s-1)\,\dot\alpha}\right\}$ and a scalar field $\Phi$ as required by higher-spin symmetry. In addition, all higher-spin fields can take values in some Lie algebra $\mathfrak{g}$.\

%%%%%%%%%%%%%%%%%%%%%%%%%%%%%%%%%%%%%%%%%%%%%%%%%%%%%%%%%%%%%%%%%%%%%%%%%%%%%%%%%
\subsection{Weinberg's soft theorem}
The universality of infrared (IR) physics in scattering processes, which captures the macroscopic dynamics of soft emitting particles, has been shown to be a rich source for uncovering hidden symmetry, structure and new physics of the $S$-matrix \cite{Cachazo:2014fwa}. One profound feature in this line of research, which dates far back to the early 1930s \cite{Bloch:1937pw}, is that it does not require a Lagrangian description. All we need is Lorentz invariance (gauge invariance) of the $S$-matrix and the existence of the soft limit, where the momentum of the emitting particle can be sent to zero. Following these criteria, Weinberg came up with an elegant theorem that is now named after him \cite{Weinberg:1964ew}. It captures the leading contribution to the $S$-matrix from the soft emission of massless particles.

\medskip

Let all massless higher-spin fields be Lorentz-real, and assume that all interactions are minimal/Noether couplings:
\begin{align}
    S_{\text{int}}=\int d^4x J_{\mu(s)}A^{\mu(s)}\,,\qquad J_{\mu(s)}=\bar{\phi}\p_{\mu_1}\ldots \p_{\mu_s}\phi+\ldots\,,
\end{align}
where $J_{\mu(s)}$ is a higher-spin conserved tensor, i.e. $\p^{\nu}J_{\nu\mu(s-1)}=0$, built out of complex scalar fields. If the emitting massless particle of helicity $h$ has real momentum $k^{\mu}$, and the $i$th external leg has momentum $p_i^{\mu}$, the current $J_i^{\mu(s)}$ where $s=|h|$ is proportional to
\begin{align}
    J_i^{\mu(s)}\sim \mathtt{c}^i_{s_i,s}\, p^{\mu_1}_i\ldots p^{\mu_{s}}_i\,,
\end{align}
in the soft limit $k^{\mu}\rightarrow 0$, where $\mathtt{c}^i_{s_i,s}$ are some coupling constants. As a consequence of the above, Poincar\'e invariance of the $n$-point scattering amplitude imposes \cite{Weinberg:1964ew}
\begin{align}
    \sum_{i=1}^n \boldsymbol{\mathtt{c}}_{s_i,s}^i\,p^{\mu_1}_i\ldots p_i^{\mu_{s-1}}=0\,.
\end{align}
For $s=1$, we recover the charge conservation law, i.e. $\sum_i\boldsymbol{\mathtt{c}}^i_{s_i,1}=0\,.$ For $s=2$, we obtain the equivalence principle since Poincare invariance plus non-triviality of $S$-matrix requires $\boldsymbol{\mathtt{c}}^i_{s_i,2}=const$, i.e. low energy graviton couples in the same way to all spins. For $s>2$, one finds the only non-trivial solution as the permutations of momenta with all coupling constants are the same. However, in a more general setting, we end up with a trivial solution: $\boldsymbol{\mathtt{c}}^i_{s_i,s\geq 3}=0$. Thus, Weinberg's soft theorem implies triviality of higher-spin $S$-matrices.\footnote{Before the class of quasi-chiral higher-spin theories was discovered, all known local theories of higher spins were shown to have vanishing amplitudes, see e.g. \cite{Joung:2015eny,Beccaria:2016syk,Roiban:2017iqg,Skvortsov:2018jea,Skvortsov:2020wtf,Skvortsov:2020gpn}.}

%Note that unlike Coleman-Mandula's theorem \cite{Coleman:1967ad} and the constructibility of higher-spin $S$-matrices using BCFW recursion \cite{Benincasa:2007xk,Benincasa:2011kn,Benincasa:2011pg}, Weinberg's soft theorem is insensitive to higher-spin symmetry and is limited to the definition of minimal/Noether couplings when considering soft higher-spin emission. For this reason, it took significant effort to discover some reasonable local higher-spin theories as discussed in Section \ref{sec:1}. %The trick, of course, is to use the chiral representation and give up parity-invariance.

%%%%%%%%%%%%%%%%%%%%%%%%%%%%%%%%%%%%%%%%%%%%%%%%%%%%%%%%%%%%%%%%%%%%%%%%%%%%
\section{Soft emission of higher-spin  particles}\label{sec:3}

We have briefly discussed the importance of choosing the `correct' field representation, i.e. the chiral one, to construct local higher-spin theories. Thus, it is natural to question how this representation can influence the conclusions of Weinberg's soft theorem. 

\medskip

In this section, we reveal the connection between IR physics/conservation laws and the number of derivatives in interacting vertices when working with the chiral representation. In particular, we observe that except for charge conservation law, all other constraints imposed by gauge invariance of the $S$-matrix are hidden in the IR. Thus, there is no restriction on having complex-valued macroscopic massless higher-spin fields. We support this observation by computing various higher-spin soft factors, which give further restriction on helicities of macroscopic massless higher-spin fields at infinity. Here, since we choose to work with the chiral representation, there will be a discrimination in treating positive and negative helicity fields. We show that conservation laws can only come from the emission of particles with positive helicities.

%%%%%%%%%%%%%%%%%%%%%%%%%%%%%%%%%%%%%%%%
\subsection{Claim}
The main result of this paper is summarized as:

\begin{thm}\label{theorem} 
Let $\cM_n(1_{h_1},...,n_{h_n})$ be an n-point non-trivial scattering amplitude of some quasi-chiral higher-spin theory where the $i$th leg has helicity $h_i$. Gauge invariance of $\cM_n$ in the soft limit implies charge conservation whenever the number of transverse derivatives in each cubic vertex is one. For vertices with a higher number of transverse derivatives, there is no constraint coming from gauge invariance of $\cM_n$ in the soft limit.\footnote{Here, we refer to the number of $\p^{0\dot 1}=\bar{\p}$ as the number of the transverse derivatives in each vertex. See terminology in \cite{Metsaev:2005ar,Ponomarev:2017nrr}.} 
\end{thm}

\proof This theorem is proved by minor propositions and results in the remainder of this section.
\qed

%%%%%%%%%%%%%%%%%%%%%%%%%%%%%%%%%%%%%%
\subsection{Initial data}\label{ID}
While $S$-matrix theory does not require a Lagrangian description a priori, it is still useful to recall some general features of higher-spin gauge potentials in the chiral representation outlined in \cite{Krasnov:2021nsq,Adamo:2022lah}.

\medskip

In 4-dimensional flat space $\M$,\footnote{Here, $\M$ can be complexified Minkowski spacetime; or another real spacetime with Euclidean or split signature.} the action for a free massless spin-$s$ higher-spin gauge potential $A_{\alpha(2s-1)\,\dot\alpha}$ has the following simple form:
\be\label{FreeGT}
S_{\mathrm{free}}=\frac{1}{2}\,\int_{\M}\partial^{\alpha}{}_{\dot\alpha}A^{\alpha(2s-1)\dot\alpha}\,\partial_{\alpha\dot\beta} A_{\alpha(2s-1)}{}^{\dot\beta}\,,\qquad \quad  \delta A_{\alpha(2s-1)\,\dot\alpha}=\partial_{\alpha\dot\alpha}\xi_{\alpha(2s-2)}
\ee
for $s\geq 1$. Upon imposing a Lorenz gauge condition of the form:
\be\label{LorG}
\partial^{\gamma\dot\alpha}A_{\alpha(2s-2)\gamma\dot\alpha}=0\,,
\ee
one can check that each higher-spin gauge potential $A_{\alpha(2s-1)\,\dot\alpha}$ contains precisely two on-shell degrees of freedom (cf., \cite{Kaparulin:2012px}). As such, we can label higher-spin fields by their helicity. 
This is an advantage of the chiral representation despite the asymmetry in determining the positive and negative helicity states.

\medskip

We assign positive helicity $+s$ to a gauge potential $A^{(+)}_{\alpha(2s-1)\,\dot\alpha}$ whenever
\be\label{ph1}
 \partial_{\alpha}{}^{\dot\gamma}A^{(+)}_{\alpha(2s-1)\,\dot\gamma}=0\,.
\ee
On the other hand, $A_{\alpha(2s-1)\,\dot\alpha}^{(-)}$ is a negative helicity $-s$ field if its curvature $F^{(-)}_{\alpha(2s)}$ obeys:
\be\label{nh1}
\partial^{\alpha\dot\alpha}F^{(-)}_{\alpha\beta(2s-1)}=0\,,\qquad \qquad  F^{(-)}_{\alpha(2s)}:=\partial_{\alpha}{}^{\dot\gamma}A^{(-)}_{\alpha(2s-1)\,\dot\gamma}\,.
\ee

Let $k^{\alpha\dot\alpha}=\kappa^{\alpha}\tilde{\kappa}^{\dot\alpha}$ be an on-shell, massless (complex) 4-momentum. We associate to $k^{\alpha\dot\alpha}$ the following on-shell positive and negative helicity polarization tensors:
\begin{align}\label{YMhel}
    \eps^{(+)}_{\alpha(2s-1)\dot\alpha}=\frac{\zeta_{\alpha(2s-1)}\tilde\kappa_{\dot\alpha}}{\kappa^{\alpha(2s-1)}\zeta_{\alpha(2s-1)}}=\frac{\zeta_{\alpha_1}\ldots \zeta_{\alpha_{2s-1}}\,\tilde{\kappa}_{\dot\alpha}}{\langle\kappa\,\zeta\rangle^{2s-1}}\,,\qquad  \eps^{(-)}_{\alpha(2s-1)\dot\alpha}&=\frac{\kappa_{\alpha_1}\ldots\kappa_{\alpha_{2s-1}}\,\tilde{\zeta}_{\dot\alpha}}{[\tilde{\kappa}\,\tilde{\zeta}]}\,,
\end{align}
where $\zeta_{\alpha},\tilde{\zeta}_{\dot\alpha}$ are constant spinors. Customarily, the notation $v_{\alpha(s-1)}$ means $v_{(\alpha_1}\ldots v_{\alpha_{s-1})}$ etc. It is easy to check that $\eps^{(+)}_{\alpha(2s-1)\,\dot\alpha}\eps_{(-)}^{\alpha(2s-1)\,\dot\alpha}=-1$. The propagator between positive and negative helicity fields in the Lorenz gauge \eqref{LorG} reads 
\begin{align}\label{propagator}
    \langle A^{(+)}_{\alpha(2s-1)\,\dot\alpha}(p)A_{(-)}^{\beta(2s'-1)\,\dot\beta}(p')\rangle = \delta^4(p+p')\tilde\delta_{s,s'} \frac{\delta_{(\alpha_1}{}^{(\beta_1}\ldots \delta_{\alpha_{2s-1})}{}^{\beta_{2s'-1})}\delta_{\dot\alpha}{}^{\dot\beta}}{p^2}\,,
\end{align}
where $\tilde\delta$ is a Kronecker delta:
\begin{equation}
    \tilde\delta(x)=\begin{cases} 0\,,\qquad x\neq 0\,,\\
    1\,,\qquad x=0\,,
    \end{cases}
\end{equation}
and we keep the spins arbitrary (for now).

%Based on the polarization configurations \eqref{YMhel}, we have 

\begin{figure}[h!]
    \centering
    \includegraphics[scale=0.21]{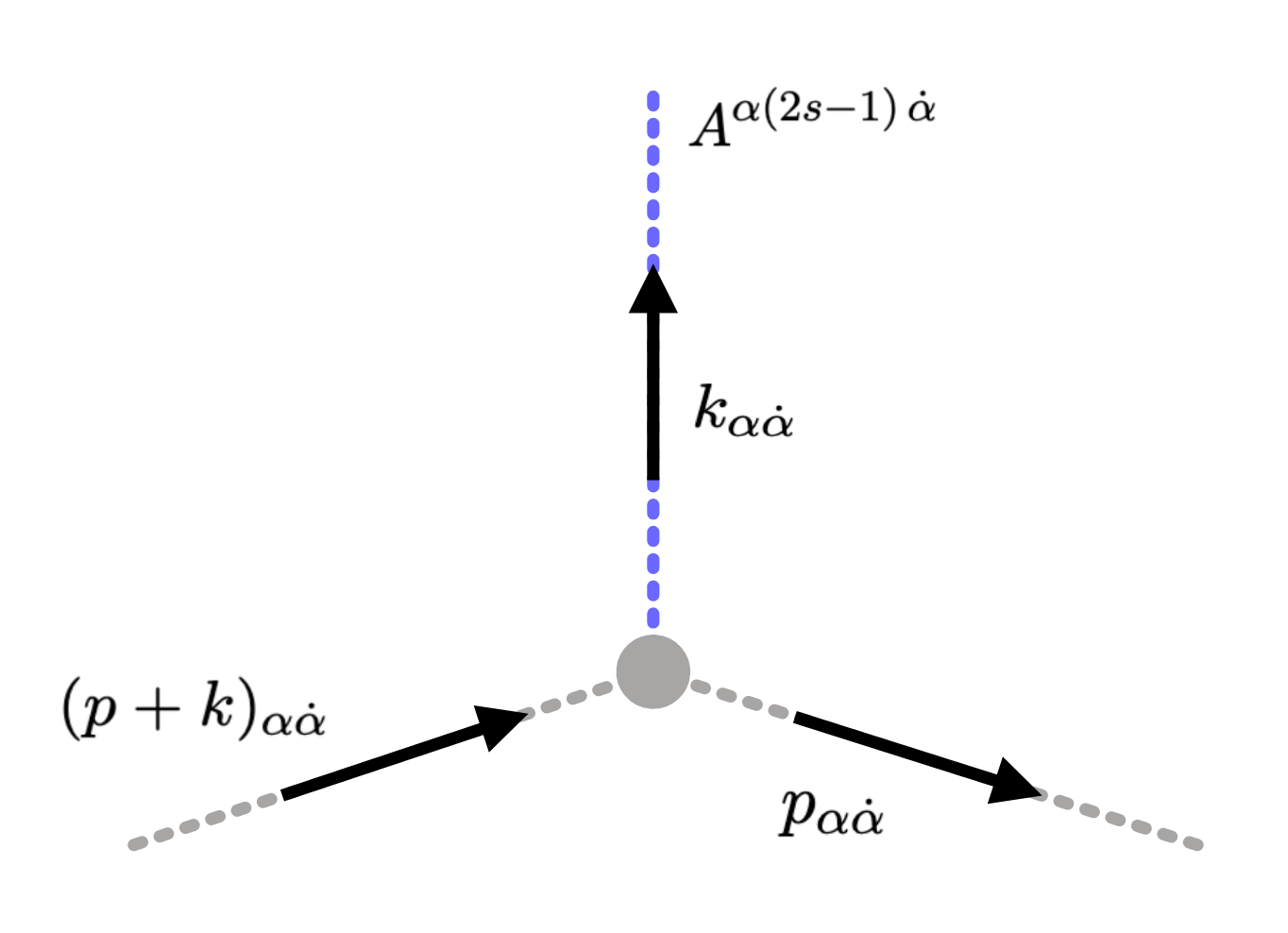}
    \caption{Soft emission of a massless higher-spin field of helicity $\pm s$ (blue) from a cubic vertex where the external leg has momentum $p_{\mu}$.}
\end{figure}

To proceed, we will work directly with the covariant cubic vertices for massless higher-spin fields derived in  \cite{Tran:2022tft}.\footnote{See also \cite{Sharapov:2022faa,Sharapov:2022awp}.} Henceforth, the momentum of the soft emitting particle is denoted as $k_{\alpha\dot\alpha}=\kappa_{\alpha}\tilde\kappa_{\dot\alpha}$ while the momentum of each external leg will be $p_i^{\alpha\dot\alpha}=\rho_i^{\alpha}\tilde\rho_i^{\dot\alpha}$. We will also assume the soft momentum can be written as
\begin{align}
    k_{\alpha\dot\alpha}=\kappa_{\alpha}\tilde\kappa_{\dot\alpha}\,,
\end{align}
The soft limit is defined by letting both types of spinors be multiplied by $\sqrt\varepsilon$
\begin{align}\label{softspinors}
    \kappa_{\alpha}\rightarrow \sqrt\varepsilon \kappa_{\alpha}\,,\qquad \tilde\kappa_{\dot\alpha}\rightarrow \sqrt{\varepsilon}\tilde\kappa_{\dot\alpha}\,,
\end{align}
where $\varepsilon$ is a small parameter. 

\medskip

There are two cases we will consider in this note: (i) emission of a massless higher-spin particle from external legs that are massless scalar fields; (ii) emission of a massless particle from external legs that are also massles higher-spin fields.

%%%%%%%%%%%%%%%%%%%%%%%%%%%%%%%%%%%%%
%%%%%%%%%%%%%%%%%%%%%%%%%%%%%%%%%%%%%%%%
\subsection{Soft emission from massless scalar fields}\label{sec:scalar}

Suppose $\cM_n(1_{\phi},\ldots,n_{\phi})$ is an $n$-point amplitude where all external legs are scalars with momentum $p_i^{\alpha\dot\alpha}$. This is the case that has the closest setup to \cite{Weinberg:1964ew}.\footnote{It should be emphasized, however, that the scalar fields considered in \cite{Weinberg:1964ew} can have mass.} The cubic vertex between two massless scalar fields and a massless higher-spin field reads \cite{Tran:2022tft}:
\begin{equation}\label{V3scalars}
    \begin{split}
    \cV_3^{0,s,0}&=\sum_m\frac{\tg_{m,i}}{m!}\,\partial_{\alpha\dot\alpha}\phi_i \underbrace{\partial_{\alpha\dot\gamma}\ldots\partial_{\alpha\dot\gamma}}_{m\,\text{times}}A^{\alpha(2m+1)\,\dot\alpha}\,\partial_{\alpha}{}^{\dot\gamma}\ldots \partial_{\alpha}{}^{\dot\gamma}\phi_i\,,\qquad m\in \mathbb{Z}_{\geq 0}\,\\
    &=\sum_m\frac{\tg_{m,i}}{m!}A^{\alpha(2m+1)\,\dot\alpha}\partial_{\alpha\dot\gamma}\ldots \partial_{\alpha\dot\gamma}\partial_{\alpha\dot\alpha}\phi_i\,\partial_{\alpha}{}^{\dot\gamma}\ldots \partial_{\alpha}{}^{\dot\gamma}\phi_i\,,
    \end{split}
\end{equation}
where we have absorbed the factor of $(-)^m$ resulting from integration by parts to the coupling constants $\tg_{m,i}$.\footnote{Note that $\partial_{\alpha\dot\gamma}\partial_{\alpha}{}^{\dot\gamma}\sim \Box \eps_{\alpha\alpha}=0$ by symmetry.}

\begin{proposition}\label{HSscalar} Gauge invariance of $\cM_n(1_{\phi},\ldots,n_{\phi})$ in the presence of a soft emitting massless higher-spin field with helicity $+m$ imposes
\begin{align}\label{claim1}
   \sum_m \varepsilon^{m}\sum_i^n 
  \tg_{m,i}\,\rho_{i\alpha(m)}\kappa_{\alpha(m)}\,[i\,\kappa]^m=0\,,\qquad m\in \Z_{\geq 0}\,.
\end{align}
 
\end{proposition}

\proof Since $\delta A^{\alpha(2m+1)\dot\alpha}=\partial^{\alpha\dot\alpha}\xi^{\alpha(2m)}$, we obtain
\begin{align}
    \delta \cV_3^{0,s,0}\sim\sum_m\frac{\tg_{m,i}}{m!}\langle i\,\kappa\rangle[\kappa\,i](p_i+k)_{\alpha\dot\gamma}\ldots (p_i+k)_{\alpha\dot\gamma}\,(p_i)_{\alpha}{}^{\dot\gamma}\ldots (p_i)_{\alpha}{}^{\dot\gamma}\,
\end{align}
in momentum space.\footnote{For convenience, we will always suppress the overall momentum conserving delta function hereafter.} Plugging in the propagator $1/\langle i\,\kappa\rangle[\kappa\,i]$, and summing over all external particles, we obtain \eqref{claim1}. 
\qed

\medskip

Observe that in the strict soft limit, where $k^{\alpha\dot\alpha},\,\kappa^{\alpha},\,\tilde\kappa^{\dot\alpha}\rightarrow 0$, the only conservation law we can get is charge conservation when $m=0$. Nevertheless, for higher-order in $\kappa$ and $\tilde\kappa$, we obtain the classical result of Weinberg which is
\begin{align}
    \sum_i^n 
  \tg_{m,i}\,p_i^{\mu_1}\ldots p_i^{\mu_m}=0\,.
\end{align}
At $m=1$, we obtain the analog of the equivalence principle as in \cite{Weinberg:1964ew}.

%%%%%%%%%%%%%%%%%%%%%%%%%%%%%%%%%%%%%%%
\subsection{Soft emission from massless higher-spin fields}\label{sec:HS}

The results of the previous subsection prompt us to answer the question: if the external fields are also massless higher-spin fields, would that change the conclusion of \cite{Weinberg:1964ew}? 

\medskip
For simplicity, we will not consider the vertices between higher-spin gauge potentials and scalar particles in this subsection. Instead, let us consider the following interactions between massless higher-spin fields \cite{Tran:2022tft}:
\begin{subequations}
\begin{align}
    \cV_3^{(+,+,+)}&=\sum_{s_i}\mathtt{C}_{h_i}^{+++}\,A^{\alpha(2|h_1|-1)}{}_{\dot\gamma}[\![A^{\alpha(2|h_2|-1)\,\dot\beta},\p_{\alpha}{}^{\dot\gamma}A^{\alpha(2|h_3|-1)}{}_{\dot\beta}]\!]\,, \label{V3plus}\\
    \cV_3^{(-,\pm,+)}&=\sum_{s_i}
    \mathtt{C}_{h_i}^{-\pm+}\,\p_{\alpha}{}^{\dot\alpha}A_{\alpha(2|h_1|-1)\,\dot\alpha}[\![A^{\alpha(2|h_2|-1)\,\dot\gamma},A^{\alpha(2|h_3|-1)}{}_{\dot\gamma}]\!]\,,\label{V3}
\end{align}
\end{subequations}
where $\mathtt{C}_{h_i}^{+++}$ and $\mathtt{C}_{h_i}^{-\pm+}$ are some dimensionful coupling constants. In addition, each gauge potential $A_{\alpha(2s-1)\,\dot\alpha}$ can carry either positive or negative helicity.\footnote{Note that the vertices $\cV_3^{(-,\pm,+)}$ can be obtained by adding $\sum_s\int B_{\alpha(2s)}B^{\alpha(2s)}$ terms to the BF action in \cite{Tran:2022tft} and intergrating out the auxiliary $B_{\alpha(2s)}$ fields.} Note that the double-square bracket $[\![\,,]\!]$ is defined as:
\begin{align}\label{partialD}
    [\![A^{\alpha(2|h_2|-1)\,\dot\gamma},A^{\alpha(2|h_3|-1)}{}_{\dot\gamma}]\!]
    :=\mathtt{f}^{\mathtt{abc}}\underbrace{\p_{\alpha\dot\beta}\ldots\p_{\alpha\dot\beta}}_{m\ \text{times}}A_{\mathtt{b}}^{\alpha(2|h_2|-1)\,\dot\gamma}\,\underbrace{\p_{\alpha}{}^{\dot\beta}\ldots\p_{\alpha}{}^{\dot\beta}}_{m\ \text{times}}A^{\alpha(2|h_3|-1)}_{\mathtt{c}\qquad \quad \   \dot\gamma}\,,
\end{align}
where $\mathtt{f}^{\mathtt{abc}}$ are the structure constants of the gauge group, and all un-dotted indices in the partial derivatives in \eqref{partialD} are contracted to those of the gauge potentials in all possible ways.\footnote{We can not help but mention that this contraction of indices coming from Moyal-Weyl deformation of twistor geometry \cite{Haehnel:2016mlb,Adamo:2016ple,Tran:2022tft} has also been discovered in the context of celestial amplitudes \cite{Ren:2022sws,Monteiro:2022lwm}.} For all un-dotted indices in the all-plus cubic vertex \eqref{V3plus} to be contracted properly, we must have $m=|h_2|+|h_3|+|h_1|-1$. While in the case of the cubic vertex $\cV_3^{(-,\pm,+)}$, it is necessary that $m=|h_2|+|h_3|-|h_1|-1$ where $|h_2|+|h_3|>|h_1|$. The dimensionful coupling constants $\mathtt{C}_{h_i}$ scales as $\mathtt{C}_{h_i}\sim\ell_p^m$ where $\ell_p$ is some natural length scale. 

\medskip

One can check that the double-square bracket produces precisely $m$ transverse derivatives $\p^{0\dot 1}=\bar{\p}$ in the light-cone gauge if the physical component of $A^{\alpha(2s-1)\,\dot\alpha}$ is $A^{1(2s-1)\,\dot 0}$. Hence, we need to insert $\ell_p^m$ so that the cubic vertex has the correct dimension. Note that $\p^{0\dot 0}=\p^+$ can be invertible in momentum space, which allows one to construct local cubic vertices in the light-cone gauge \cite{Metsaev:2005ar}. Note that the amplitudes resulting from the above vertices can be found in \eqref{allplus}, \eqref{MHV-bar0} and \eqref{MHV}.

\medskip

To relate to Weinberg's arguments, it is necessary to have a maximal number of external momentum $p_i^{\alpha\dot\alpha}$ in the interactions. Thus, without loss of generalization, we will assume that the soft emitting particle will always be the first leg with helicity $h_1=\pm s$ while the remaining will be the external leg $i$th with momentum $p_i^{\alpha\dot\alpha}$ and the internal propagator with momentum $(p_i+k)^{\alpha\dot\alpha}$. 

\medskip

To proceed, we shall fix the spin of the soft particle to be $s$ and the number of derivatives to be $m$ for all couplings. We will also write all coupling constants in terms of $s,m$ and the label $i$ of external legs as:
\begin{align}
    \tg^{+++}_{s,m,i}\,,\qquad \tg^{-\pm +}_{s,m,i}\,,
\end{align}
to match with the pattern of \eqref{mainresult}. The explicit form of $\tg$ will not be important in the following discussions. There are two scenarios to be investigated separately:

\medskip

$\diamond$ \underline{\textbf{Scenario I:} The soft particle is emitted from all-plus vertices.} 

As noted in \cite{Tran:2022tft}, there must be at least one extra pair of derivatives for \eqref{V3plus} to make sense. Furthermore, \eqref{V3plus} represents non-minimal couplings since it has the maximal number of derivatives allowed by kinematics. To be more concrete, consider the case where all fields are spin-1 fields. For all un-dotted indices in \eqref{V3plus} to be contracted properly, there must be three transverse derivatives. This is obviously different to the usual (minimal) gauge interaction with only one transverse derivative.

\begin{proposition}\label{prop1} In the presence of a soft emitting massless higher-spin particle from all-plus vertices, gauge invariance of $\cM_n$ is trivially satisfied in the soft limit.
\end{proposition}

\proof Under a gauge transformation in momentum space, $\delta \cV_3^{(+,+,+)}$ results in 
\begin{align}\label{step1}
    0=\delta \cV_3^{(+,+,+)}\sim\sum_{m}\tg_{s,m,i}^{+++}\langle i\,\kappa\rangle[\kappa\,i] \rho_{i\alpha(m)}\kappa_{\alpha(m)}[i\,\kappa]^m\,,\qquad m\geq 1\,.
\end{align}
While the denominator coming from the propagator can remove the factor $\langle i\,\kappa\rangle[\kappa\,i]$ in \eqref{step1}, it is obvious that the above is trivially satisfied in the soft limit if $m\geq 1$. 
\qed

\medskip

Proposition \ref{prop1} implies that non-minimal couplings do not provide new physics for soft higher-spin scatterings. This is indeed in agreement with previous discussion in \cite{Damour:1987fp,Bekaert:2010hw}. Here, we once again obtain the series of constraints in \eqref{mainresult} for $\forall m\geq 1$.

\medskip

$\diamond$ \underline{\textbf{Scenario II:} The soft particle is emitted from $\cV_3^{(-,\pm,+)}$ vertices.}

\begin{proposition} Gauge invariance of $\cM_n(1_{h_1},\ldots,n_{h_n})$ in the presence of a soft emitting massless higher-spin field with helicity $\pm s$ imposes
\begin{align}\label{claim2}
    \varepsilon^{m}\sum_i^n \tg^{-\pm +}_{s,m,i}\rho_i^{\alpha(m)}\kappa^{\alpha(m)}[i\,\kappa]^m=0\,, \qquad m\in\Z_{\geq 0}\,.
\end{align}

\end{proposition}

\proof We rewrite \eqref{V3} as $\partial_{\alpha}{}^{\dot\alpha}A_{\alpha(2|h_1|-1)\,\dot\gamma}[\![A^{\alpha(2|h_2|-1)\,\dot\gamma},A^{\alpha(2|s_3|-1)}{}_{\dot\alpha}]\!]$ by symmetrizing over external legs. Then, by proceeding similarly to Proposition \ref{prop1}, we arrive at \eqref{claim2}.
\qed

%%%%%%%%%%
\medskip

Obviously, when $m=0$, \eqref{claim2} reduces to the usual charge conservation. At $m=1$, we recover the equivalence principle where $\tg^{-++}_{s,1,i}=const$. Intriguingly, all constraints imposed by gauge invariance are `hidden' in the IR when we work with the chiral representation. 

%%%%%%%%%%%%%%%%%%%%%%%%%%%%%%%%%%%%%%%%%

\subsection{Higher-spin soft factors}
In the analysis of the previous subsections, we do not know how the conservation laws are related to the helicity of the soft particle. Therefore, to find further constraints on macroscopic massless higher-spin fields, we need to compute soft factors associated to \eqref{V3scalars}, \eqref{V3plus} and \eqref{V3}. Since (quasi-)chiral theories are local, we can employ BCFW recursion techniques \cite{Britto:2005fq} to study the factorization of tree-level amplitudes in the soft limit. We discover from our computation that conservation laws can only come from soft emitting particles with positive helicity.

\medskip

Suppose $\cM_{n+1}(k_{\pm s},1_{h_1},\ldots,n_{h_n})$ is an $(n+1)$-point tree-level scattering amplitude where $k$ is the soft particle with helicity $\pm s$ and momentum $k_{\alpha\dot\alpha}$. For $\cM_{n+1}$ to be analytic, its simple poles must come from the exchange propagators, and it must decay sufficiently fast when the value of the deform parameter $z$ is large.

\begin{figure}[ht]
    \centering
    \includegraphics[scale=0.37]{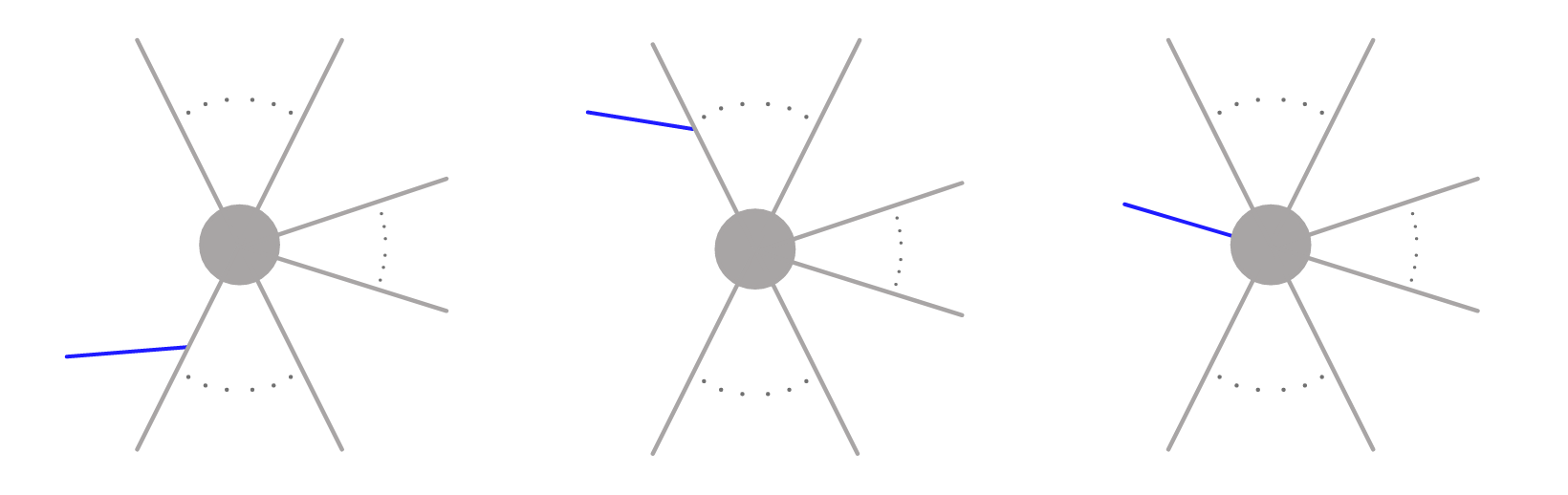}
    \caption{Contributions to soft higher-spin emission amplitude where the emitting particle is a complex-valued higher-spin field (blue).}
\end{figure}

Without loss of generality, we assume that the $n$th particle always has negative helicity in the case it is a spinning field. The spinors associated with this external leg, whether chiral or anti-chiral, will be considered as reference spinors. Under these assumptions, the $(n+1)$-point amplitude $\cM_{n+1}$ factorizes (schematically) as
\begin{align}
    \cM_{n+1}=\sum_a\cM_L(k(z^*),1,\ldots,a,P_I)\frac{1}{P_I^2}\cM_R(-P_I,a+1,\ldots,n(z^*))\,,
\end{align}
where $z^*$ is the location of the pole in the denominator $P_I^2=(k+\sum_{a\in I}p_a)^2=0$ for any non-empty subset $I=\{1,\ldots,a\}\subset \{1,\ldots,n-1\}$. Since we are only interested in soft emission of massless higher-spin particles, it is sufficient to focus on the case where $\cM_L$ are 3-pt amplitudes \cite{Cachazo:2014fwa}:
\begin{align}\label{factorization3pt}
    \cM_3(k(z^*),i_{h_i},P_I)\frac{1}{P_I^2}\cM_n(-P_I,2_{h_2},\ldots, n_{h_n}(z^*))\,.
\end{align}
Note that to have non-trivial higher-spin couplings, $\cM_3$ must be non-zero and non-singular. In order to obtain conservation laws from \eqref{factorization3pt}, we will need to sum over all particles and extract IR physics from the soft limit of the factorization channels:\footnote{Note that for chiral higher-spin theories, all factorization channels are trivial at tree-level.}
\begin{align}\label{factorizedchannels}
    \sum_{i=1}^{n-1}\cM_3(k(z^*),i_{h_i},P_I)\frac{1}{P_I^2}\cM_n(-P_I,2_{h_2},\ldots, n_{h_n}(z^*))\,.
\end{align}

One can check that $\cM_{3}(-,-,-)$ vanishes on-shell regardless of whether we consider \eqref{V3plus} or \eqref{V3}. Hence, all 3-point amplitudes where external legs are spinning fields we can have are $\cM_3(+,+,+)$, $\cM_3(-,+,+)$ or $\cM_3(-,-,+)$.  

\medskip

\paragraph{Building blocks.} To have a non-trivial $\cM_3(0,0,s)$ amplitude from the vertex \eqref{V3scalars}, the potential $A^{\alpha(2m+1)\,\dot\alpha}$ must carry positive helicity. A short computation gives us the following 3-pt amplitude
\begin{align}\label{3pt00s}
    \cM_3(1_0,2_{+m},3_{0})\sim \frac{[1\,2]^{m+1}[2\,3]^{m+1}}{[3\,1]^{m+1}}\,,
\end{align}
where we have utilized
\begin{align}
    \langle \zeta\, 1\rangle [1\,3]+\langle \zeta\, 2\rangle [2\,3]=0\,,\qquad \qquad \langle \zeta\, 1\rangle [1\,2]+\langle \zeta\, 3\rangle [3\,2]=0\,,
\end{align}
on the support of (complex) momentum conservation. 

\medskip

Next, using the polarizations in \eqref{YMhel} and complex on-shell momenta $p_{i}^{\alpha\dot\alpha}=\rho_{i}^{\alpha}\tilde{\rho}_{i}^{\dot\alpha}$, where $\tilde{\rho}_{i}^{\dot\alpha}\neq \overline{\rho_i^{\dot\alpha}}$, we obtain the 3-point all-plus helicity amplitude as
\begin{equation}\label{allplus}
\begin{split}
    \cM_3(1_{s_1}^+,2_{s_2}^+,3_{s_3}^+)&=\tilde\delta(m+2-(s_1+s_2+s_3))[12]^{s_1+s_2-s_3}[23]^{s_2+s_3-s_1}[31]^{s_3+s_1-s_2}\,\\
    &=\tilde\delta(m+2-(s_1+s_2+s_3))\Big(\frac{[23][31]}{[12]}\Big)^{m+2}\Big(\frac{[12]}{[23]}\Big)^{2s_1}\Big(\frac{[12]}{[31]}\Big)^{2s_2}\,.
\end{split}
\end{equation}

Moving on, the $\overline{\text{MHV}}_3$ amplitudes read:\footnote{All the prefactors will not play a significant role in this work as we only focus on constructing the higher-spin soft factors in the chiral representation.}
\be\label{MHV-bar0}
    \begin{split}
\cM_{3}(1_{s_1}^-,2_{s_2}^+,3_{s_3}^+)&\sim\,\tilde{\delta}(m+1-(s_2+s_3-s_1))\,\frac{[23]^{2s_2+2s_3-m-1}}{[31]^{2s_2-m-1}[12]^{2s_3-m-1}}\\
&\sim \tilde\delta(m+1-(s_2+s_3-s_1))\Big(\frac{[12][31]}{[23]}\Big)^m\frac{[23]^{2s_2+2s_3-1}}{[31]^{2s_2-1}[12]^{2s_3-1}}\,.
\end{split}
\ee

\medskip

Computing the MHV${}_3$ amplitudes requires a symmetrization over the positions of two negative helicity external fields while keeping the location of the positive helicity particle intact. A simple computation shows that
\be\label{MHV}
    \begin{split}
\cM_{3}(1_{s_1}^{-},2_{s_2}^{-},3_{s_3}^{+})&\sim\tilde\delta(m+1-(s_2+s_3-s_1))\frac{1}{2}\frac{\langle 12\rangle^{2(s_2+s_3)-m-1}[31]^m}{\langle 23\rangle^{2s_3-m-1}\langle 31\rangle }+(1\leftrightarrow 2)\\
&\sim\tilde\delta(m+1-(s_2+s_3-s_1))\frac{1}{2}\frac{\langle 12\rangle^{2(s_2+s_3)-1}}{\langle 23\rangle^{2s_3-1}\langle 31\rangle}\Big(\frac{[12][31]}{[23]}\Big)^m+(1\leftrightarrow 2)\,.
\end{split}
\ee
Notice that when $m=0$, \eqref{MHV} reduces to the MHV amplitude of HS-YM theory \cite{Adamo:2022lah}. Furthermore, if all external spins are $s_1=s_2=s_3=1$, there will be no higher-derivative interaction since $m$ is forced to be zero on the support of the Kronecker delta $\tilde \delta$. This demonstrates the deep connection between all quasi-chiral higher-spin theories and the action functional of YM theory obtained by Chalmers and Siegel~\cite{Chalmers:1996rq}. Lastly, it is intriguing to note that the $\overline{\text{MHV}}_3$ amplitude is not the helicity conjugate of the MHV${}_3$ one for generic values of spins and number of derivatives. This is an inevitable consequence when working with fields in the chiral representation (cf., \eqref{YMhel}).

\medskip

To proceed, let us compute the 4-pt MHV amplitudes $\cM_4(1_{+s_1},2_{+s_2},3_{-s_3},4_{-s_4})$ from \eqref{MHV-bar0} and \eqref{MHV} using BCFW recursion. By solving all the constraints from the Kronecker deltas under the requirement that the amplitudes should have well-behaviour factorization (see discussion in \cite{Adamo:2022lah}), we obtain:
\be\label{4pt1}
    \begin{split}
  \cM_4(1_{+s_1},&2_{+s_2},3_{-s_3},4_{-s_4})=\sum_{m=0}^{\infty}\mathtt{C}_{h_1,h_2,m}\mathtt{C}_{m,h_3,h_4}\tilde\delta(s_3-s_4)\\
  &\times\frac{[1\,2]^{2s_1+2s_2-1}}{[2\,\hat P]^{2s_1-1}[\hat P\,1]^{2s_2-1}}\Big(\frac{[\hat P\,\hat 1][2\,\hat P]}{[\hat 1\,2]}\Big)^{s_1+s_2-m-2}\frac{\langle 3\,4\rangle^{2(s_3+m)+1}}{\langle \hat 4\,\hat P\rangle^{2m+1}\langle \hat P\,3\rangle}\Big(\frac{[3\,\hat 4][\hat P\,3]}{[4\,\hat P]}\Big)^{m}\\
  &\qquad +(3\leftrightarrow 4)\,.
  \end{split}
\ee
for some couplings $\mathtt{C}$; and we have performed the BCFW $[-+\rangle$-shift for which
\begin{subequations}
\begin{align}
    \rho_1^{\alpha}\rightarrow \hat\rho_1^{\alpha}(z):&=\rho_1^{\alpha}+z\,\rho_4^{\alpha}\,,\\
    \tilde\rho_4^{\dot\alpha}\rightarrow \hat{\tilde\rho}^{\dot\alpha}_4(z):&=\tilde\rho_4^{\dot\alpha}-z\,\tilde\rho_1^{\dot\alpha}\,,\\
    \hat P^{\alpha\dot\alpha}(z):&=\rho_1^{\alpha}\tilde\rho_1^{\dot\alpha}+\rho_2^{\alpha}\tilde\rho_2^{\dot\alpha}+z\,\rho_4^{\alpha}\tilde\rho_1^{\dot\alpha}\,,\\
    z:&=\frac{\langle 1\,2\rangle}{\langle 2\,4\rangle}\,.
\end{align}
\end{subequations}
Observes that unless $m=0$, we will have trivial amplitudes since {\small $[3\,\hat 4]=[3\,4]-\frac{\langle 1\,2\rangle}{\langle2\,4\rangle}[3\,1]=0$}.\footnote{This implies that the vertices \eqref{V3} cannot be used to construct quasi-chiral theories with higher-derivative interactions. However, it does not mean quasi-chiral theories with higher-derivative interactions cannot exist.} As a result, $\cM_4(1_{+s_1},2_{+s_2},3_{-s_3},4_{-s_4})$ reduces to the 4-pt MHV amplitude of HS-YM, which is
\begin{align}\label{4ptHSYM}
    \cM_4(1_{+1},2_{+1},3_{-s},4_{-s})\sim \frac{\langle 3\,4\rangle^{2s+2}}{\langle 1\,2\rangle\langle 2\,3\rangle\langle 3\,4\rangle\langle 4\,1\rangle}\,.
\end{align}

\paragraph{Comments on soft limit.} First of all, if the positive helicity particles (either 1 or 2) go soft under the parametrization \eqref{softspinors}, then \eqref{4ptHSYM} scales as
    \begin{align}
        \cM_4(1_{+1},2_{+1},3_{-s},4_{-s})\sim \frac{1}{\varepsilon}\frac{\langle 3\,4\rangle^{2s+2}}{\langle 1\,2\rangle\langle 2\,3\rangle\langle 3\,4\rangle\langle 4\,1\rangle}\,.
    \end{align}
However, if we send the momentum of the negative helicity particles (either 3 or 4) to zero, then \eqref{4ptHSYM} gets heavily suppressed since 
\begin{align}
    \cM_4(1_{+1},2_{+1},3_{-s},4_{-s})\sim \varepsilon^{s}\frac{\langle 3\,4\rangle^{2s+2}}{\langle 1\,2\rangle\langle 2\,3\rangle\langle 3\,4\rangle\langle 4\,1\rangle}\,.
\end{align}
The above indicates that soft limit of negative helicity particles can always be taken smoothly without the risk of creating divergences in the IR. 

\medskip

Note that we do not have concrete examples of N${}^{k\geq 1}$MHV amplitudes of HS-YM theory nor some non-vanishing amplitudes of a quasi-chiral higher-spin theory with higher-derivative interactions. Nevertheless, it is still reasonable to `expect' that negative helicity particles will always have a smooth soft limit while soft positive helicity particles will be responsible for all IR physics when working with the chiral representation. To wit, we can have macroscopic higher-spin fields of negative helicity but not positive ones.\footnote{Note that the soft factors can be used to fix the amplitudes, see e.g. \cite{Rodina:2018pcb}. We will explore this possibility in a future work.}

%%%%%%%%%%%%%%%%%%%%%%%%%%%%%%%%%%%%
\paragraph{Soft factors from $\cM_3(0,0,+s)$ amplitudes.} Suppose $\cM_n(1_{\phi},\ldots,n_{\phi})$ is an $n$-point amplitude where all external legs are scalar fields with momentum $p_i^{\alpha\dot\alpha}$. Since external legs are not spinning fields, we do not need to consider a BCFW shift. It is useful to parametrize the spinors of the exchange with momentum $P^{\alpha\dot\alpha}=(p_i+k)^{\alpha\dot\alpha}$ as
\begin{align}\label{internalspinor1}
    P^{\alpha}=\rho_i^{\alpha}\,,\qquad \tilde P^{\dot\alpha}=\frac{\langle \tau\,\kappa\rangle}{\langle \tau \,i \rangle}\tilde\kappa^{\dot\alpha}+\tilde\rho_i^{\dot\alpha}\,,
\end{align}
where $\tau^{\alpha}\,,\tilde\tau^{\dot\alpha}$ are some reference/constant spinors.

\begin{proposition}\label{scalar} Gauge invariance of $\cM_n(1_{\phi},\ldots,n_{\phi})$ in the presence of a soft emitting massless higher-spin field with helicity $+m$ imposes
\begin{align}\label{claim4}
    \sum_i^n \frac{\tg_{m,i}}{m!}\rho_i^{\alpha(m)}\tilde\rho_i^{\dot\alpha(m)}\tilde\kappa_{\dot\alpha(m)}=0\,.
\end{align}
\end{proposition}

\proof Substituting \eqref{internalspinor1} to \eqref{3pt00s}, we obtain
\begin{align}
    \cM_3(i_0,k_{+s},P_0)= \frac{\tg_{m,i}}{m!}\,\frac{\langle \tau\,i\rangle^{m+1}[i\,\kappa]^{m+1}}{\langle \tau\,\kappa\rangle^{m+1}}=\frac{\tg_{m,i}}{m!}\eps^{(+)}_{\alpha(m+1)\,\dot\alpha}\rho_i^{\alpha(m+1)}\tilde\rho_i^{\dot\alpha(m+1)}\tilde\kappa_{\dot\alpha(m)}\,.
\end{align}
where the polarization tensor for positive helicity field has been written as
\begin{align}
    \eps^{(+)}_{\alpha(m+1)\dot\alpha}=\frac{\tau_{\alpha(m+1)}\tilde\kappa_{\dot\alpha}}{\langle\kappa\,\tau\rangle^{m+1}}\
\end{align}
by virtue of \eqref{YMhel}. Plugging in the propagator, one gets
\begin{align}
    \cK_{m,i}^0=\frac{\tg_{m,i}}{m!}\frac{\eps^{(+)}_{\alpha(m+1)\,\dot\alpha}\rho_i^{\alpha(m+1)}\tilde\rho_i^{\dot\alpha(m+1)}\tilde\kappa_{\dot\alpha(m)}}{\langle i\,\kappa\rangle[\kappa\,i]}\,.
\end{align}
We shall refer to $\cK_{m,i}^0$ as higher-spin soft factor associated to external legs of spin-0. Since $\eps_{\alpha(m+1)\,\dot\alpha}$ transforms as $\delta \eps_{\alpha(m+1)\,\dot\alpha}=k_{\alpha\dot\alpha}\,\xi_{\alpha(m)}$ (after gauge fixing), Lorentz invariance of $\cM_n$ imposes \eqref{claim4}.
\qed

\medskip

Once again we find that when there is only one transverse derivative in each vertex, i.e. $m=0$, Poincar\'e invariance of the $S$-matrix implies charge conservation since
    \begin{align}\label{imder1}
        \sum_i^n\, \tilde\delta_{m,0}\tg_{m,i}=0\,,
    \end{align}
When $m>0$, we arrive at the same conclusion with previous subsections, i.e. there is no restriction on helicities of macroscopic higher-spin fields if the soft limit is strictly applied.

\paragraph{Soft factors from all-plus helicity amplitudes.} As the name suggests, the emitting particle can only carry positive helicity in this case. Consider the following BCFW shift:
\begin{subequations}\label{deformplus}
\begin{align}
    \kappa^{\alpha}&\rightarrow \hat\kappa^{\alpha}(z):=\kappa^{\alpha}+z\,\rho_n^{\alpha}\,,\\
    \tilde\rho_n^{\dot\alpha}&\rightarrow \hat{\tilde{\rho}}_n^{\dot\alpha}(z):=\tilde\rho_n^{\dot\alpha}-z\, \tilde\kappa^{\dot\alpha}\,,
\end{align}
\end{subequations}
where the critical value $z^*$ of the deformation parameter associated with the factorization \eqref{factorization3pt} reads
\begin{align}
    (k(z^*)+p_i)^2=0\qquad \Rightarrow \qquad z^*=\frac{\langle \kappa\, i\rangle}{\langle i\,n\rangle}\,.
\end{align}
Using the parametrization \eqref{internalspinor1} and feeding the above information back to \eqref{allplus}, we obtain the following soft factor
\be\label{cFallplus}
\begin{split}
   \cG_{s,m,i}= \frac{\cM_3(i_{+s_i},P_{+\omega},k_{+s})}{\langle i\,\kappa\rangle[\kappa\,i]}= \tg_{s,m,i}^{+++}\, \frac{[i\,\kappa]^{m+1}\langle n\,i\rangle^{2s-m-2}}{\langle i\,\kappa\rangle\langle n\,\kappa\rangle^{2s-m-2}}
    \end{split}
\ee
on the support of the Kronecker delta $\tilde\delta(m+2-(s+s_i+\omega))$. In the soft limit, $\cG_{s,m,i}$ scales as
\begin{subequations}\label{scalingofFbarplus}
\begin{align}
    \cG_{s,m,i}&\sim \frac{1}{\varepsilon^{s-m-1}}\,\frac{[i\,\kappa]^{m+1}\langle n\,i\rangle^{2s-m-2}}{\langle i\,\kappa\rangle\langle n\,\kappa\rangle^{2s-m-2}}\,.
\end{align}
\end{subequations}
Observe that the higher the spin, the more singular $\cG_{s,m,i}$ is in the soft limit. Intriguingly, we can mitigate this effect by having higher number of derivatives in the cubic vertex \eqref{V3plus}. Furthermore, it should be possible to study subleading contributions to 3-pt factorizations of tree-level higher-spin amplitudes in the soft limit as in \cite{Cachazo:2014fwa}.\footnote{We postpone this study for a future work.}

%%%%%%%%%%%%%%%%%%%%%%%%%%%%%%%%%%
\paragraph{Soft factors from $\overline{\text{MHV}}_3$ amplitudes.} There are two sub-cases to study:

\begin{enumerate}
    \item \underline{\textit{The emitting particle has helicity $+s$:}} Consider 3-pt amplitude $\cM_3(P_{-\omega},i_{+s_i},k_{+s})$:
\be
\begin{split}
\cM_3(P_{-\omega},i_{+s_i},k_{+s})\sim\,\frac{[i\,\kappa]^{m+1}\langle n\,i\rangle^{2s-m-1}}{\langle n\,\kappa\rangle^{2s-m-1}}\,,\qquad  (m\in \Z_{\geq 0})\,.
\end{split}
\ee
Unlike the case of HS-YM, where the exchange particle can only be gluon \cite{Adamo:2022lah}, here we can have higher-spin fields in the exchange due to higher-derivative interactions. Plugging in the propagator $1/\langle i\,\kappa\rangle[\kappa \,i]$, we arrive at the following higher-spin soft factors for positive helicity fields:
\begin{align}\label{softFbarplus}
    \bar{\cF}^{(+)}_{s,m,i}=\tg_{s,m,i}^{-++}\,\frac{[i\,\kappa]^{m}\langle n\,i\rangle^{2s-m-1}}{\langle i\,\kappa\rangle\langle n\,\kappa\rangle^{2s-m-1}}\,,\qquad  (m\in \Z_{\geq 0})\,.
\end{align}
%where we introduced some coupling constant $\tg_{s,i}$ for later discussion.

It is interesting to note that when $s=1,\,m=0$, \eqref{softFbarplus} reduces to the usual soft factor of gauge theory \cite{Casali:2014xpa}; and when $s=2,\,m=1$, the above soft factor reduce to the usual one of gravity \cite{Cachazo:2014fwa}. To be more concrete let us spell them out explicitly
\begin{subequations}
\begin{align}
    s&=1\,,\,m=0 \quad  (\text{gauge})  &&: &&  \cF_{1,0,i}=\tg^{-++}_{1,0,i}\frac{\langle n\,i\rangle}{\langle i\,\kappa\rangle \langle n\,\kappa\rangle}\,,\\
    s&=2\,,\,m=1 \quad  (\text{gravity}) &&: && \cF_{2,1,i}=\tg^{-++}_{2,1,i}\frac{[i\,\kappa]\langle n\,i\rangle^2}{\langle i\,\kappa\rangle \langle n\,\kappa\rangle^2}\,.
\end{align}
\end{subequations}

In the soft limit, the soft factors $\bar\cF^{(+)}_{s,m,i}$ scale as
\begin{subequations}
\begin{align}
    \bar\cF^{(+)}_{s,m,i}&\sim \frac{1}{\varepsilon^{s-m}}\,\frac{[i\,\kappa]^{m}\langle n\,i\rangle^{2s-m-1}}{\langle i\,\kappa\rangle\langle n\,\kappa\rangle^{2s-m-1}}\,.
\end{align}
\end{subequations}
Once again, we observe that singularity behaviour can be soften by having higher number of derivatives.

\item \underline{\textit{The emitting particle has helicity $-s$:}} In this case, $\cM_3(k_{-s},P_{+\omega},i_{+s_i})$ reads
\be
\begin{split}\label{MHV3case2}
\cM_{3}(k_{-s},P_{+\omega},i_{+s_i})\sim \frac{[i\,\kappa]^{m+1}\langle n\,\kappa\rangle^{2s+m+1}}{\langle n\,i\rangle^{2s+m+1}}\,.
\end{split}
\ee
Hence, 
\begin{align}
    \bar{\cF}^{(-)}_{s,m,i}\sim \frac{[i\,\kappa]^m\langle n\,\kappa\rangle^{2s+m+1}}{\langle i\,\kappa\rangle\langle n\,i\rangle^{2s+m+1}}\,.
\end{align}
It is easy to check that $\bar{\cF}^{(-)}$ is heavily suppressed in the soft limit since
\begin{align}
    \bar{\cF}^{(-)}_{s,m,i}\sim \varepsilon^{s+m}\frac{[i\,\kappa]^{m}\langle n\,\kappa\rangle^{2s+m+1}}{\langle i\,\kappa\rangle\langle n\,i\rangle^{2s+m+1}}\,.
\end{align}
Thus, the soft factor $\bar{\cF}^{(-)}_{s,m,i}$ does not give us new information on IR physics of soft higher-spin emission. This is equivalent to say that a soft particle emitted from an $\cV_3^{(-,+,+)}$ vertex is allowed to have arbitrary negative helicity as long as it obeys higher-spin symmetry.
\end{enumerate}
   
%%%%%%%%%%%%%%%%%%%%%%%%%%%%%%%%%%%
\paragraph{Soft factors from MHV${}_3$ amplitudes.} Again, there are two sub-cases:
\begin{enumerate}
    \item \underline{\textit{The emitting particle has positive helicity $+s$:}}
    
    Consider the factorized 3-pt amplitude $\cM_3(P_{-\omega},i_{-s_i},k_{+s})$. To avoid singularity, we have to set $m=0$. Using the parametrization \eqref{internalspinor1}, it is simple to show that 
    \begin{align}
        \cM_3(P_{-\omega},i_{-s_i},k_{+s})\sim 0\,,
    \end{align}
    This yields the following soft factor
    \begin{align}\label{softFplus}
        \cF^{(+)}_{s,m,i}=0\,.
    \end{align}
    
    \item \underline{\textit{The emitting particle has negative helicity $-s$:}}
    
    Once again, we have to set $m=0$ to avoid singularity and perform the following BCFW shift:
    \begin{subequations}\label{deformminus}
    \begin{align}
        \tilde\kappa^{\dot\alpha}&\rightarrow \hat{\tilde{\kappa}}^{\dot\alpha}(z):=\tilde\kappa^{\dot\alpha}-z\,\tilde\rho_n^{\dot\alpha}\,,\\
        \rho_n^{\alpha}&\rightarrow \hat{\rho}_n^{\alpha}(z):=\rho_n^{\alpha}+z\, \kappa^{\alpha}\,.
    \end{align}
    \end{subequations}
    Furthermore, we parametrize the internal spinors as\footnote{Note that there is no unique way of doing this.}
    \begin{align}\label{internalspinor2}
        P^{\alpha}=\rho_i^{\alpha}+\frac{[ n\,\kappa]}{[ n \,i ]}\kappa^{\alpha}\,,\qquad \tilde P^{\dot\alpha}=\tilde\rho_i^{\dot\alpha}
    \end{align}
    we obtain:
    \begin{align}
        \cM_3(-i_{s_i},k_{-s},P_{+\omega})\sim \,\frac{\langle i\,\kappa\rangle^{2s-1}}{2}\frac{[n\,i]}{[n\,\kappa]}\Big(1+\frac{[i\,n]^{2\omega-2}}{[n\,\kappa]^{2\omega-2}}\Big)\,,
    \end{align}
    where $\omega$ stands for the spin of the exchanges and $s$ is the spin of the soft emitting particle as usual. Here, the spin constraint imposes $s_i=s$ as in \cite{Adamo:2022lah}. The soft factor reads
    \begin{equation}\label{softFminus}
        \cF^{(-)}_{s,m,i}=\,\frac{\langle i\,\kappa\rangle^{2s-2}}{2}\frac{[n\,i]}{[\kappa\,i][n\,\kappa]}\Big(1+\frac{[i\,n]^{2\omega-2}}{[n\,\kappa]^{2\omega-2}}\Big)\,,
    \qquad  (m\in \Z_{\geq 0})\,,
    \end{equation}
    Observe that all soft factors obtained from the MHV${}_3$ amplitudes are not the helicity conjugate of \eqref{softFbarplus} for generic spin $s$ except for the case $s=\omega=1$ (gauge theory), i.e.
    \begin{align}
        \overline{\cF^{(-)}_{1,0,i}}=\bar{\cF}^{(+)}_{1,0,i}\,.
    \end{align}
In the soft limit, it is easy to see that $\cF^{(-)}_{s,m,i}$ is heavily suppressed. Thus, it is irrelevant for IR physics of soft higher-spin emission.
\end{enumerate}    
Note that all of the soft factors above do not depend on spin/helicity/momenta of the particles adjacent to the soft emitting particle which is similar to the previous examples of YM \cite{Casali:2014xpa} and GR \cite{Cachazo:2014fwa}.

%%%%%%%%%%%%%%%%%%%%%%%%%%%%%%%%%%%%%%%%%%%

%%%%%%%%%%%%%%%%%%%%%%%%%%%%%%%%%%%%%%%%
\subsection{Implications from higher-spin soft factors}
Armed with all soft factors in the previous subsection, we are now ready to discuss their `physical' implications. 

\medskip

Of all the soft factors computed above, we see that only $\cG_{s,m,i}$ and $\bar{\cF}^{(+)}_{s,m,i}$ have singular behaviour while others are suppressed in the soft limit. Thus, we only need to focus on $\cG_{s,m,i}$ and $\bar{\cF}^{(+)}_{s,m,i}$ to extract IR physics of soft higher-spin emission.

Suppose the spin of the emitting particle is $s$ and the number of derivatives in each vertex is $m$. We observe that it is convenient to combine these two numbers together to form the `effective helicity' of the emitting particle. This allows spins and derivatives to `compete' to keep the total momentum/spinors in the range where IR physics can be extracted.

\medskip

$\diamond$ \underline{\textbf{Case I:} Implication from soft limit of $\cG_{s,m,i}$.} 

For convenience, we will cast the soft factors $\cG_{s,m,i}$ into the following form:
\be\label{sceIstep1}
\begin{split}
   \cG_{s,m,i}=\tg_{s,m,i}^{+++}\, \frac{[i\,\kappa]^{m+2}\langle n\,i\rangle^{2s-m-2}}{k\cdot p_i\,\langle n\,\kappa\rangle^{2s-m-2}}=\frac{\tg_{s,m,i}^{+++}}{k\cdot p_i} \eps^{(+)}_{\alpha(2s-m-2)\,\dot\alpha}\tilde\rho_i^{\dot\alpha(m+2)}\rho_i^{\alpha(2s-m-2)}\tilde\kappa_{\dot\alpha(m+1)}\,.
    \end{split}
\ee
Here, we have treated $\rho_n^{\alpha}\,,\tilde\rho_n^{\dot\alpha}$ as reference spinors and written the effective polarization tensor as
\begin{align}
    \eps^{(+)}_{\alpha(2s-m-2)\dot\alpha}=\frac{\rho_{n\,\alpha(2s-m-2)}\tilde\kappa_{\dot\alpha}}{\langle\kappa\,n\rangle^{2s-m-2}}\,.
\end{align}

\begin{proposition}\label{prop2} Gauge invariance of an $n$-point scattering amplitude $\cM_n$ when there is a soft emitting particle with positive helicity field $+s$ imposes:
\begin{align}\label{claim5}
    \sum_i \tg_{s,m,i}^{+++}\,\rho_i^{\alpha(2s-m-3)}\tilde\rho_i^{\dot\alpha(m+1)}\tilde\kappa_{\dot\alpha(m+1)}=0\,.
\end{align}
\end{proposition}

\proof Let $\eps_{\alpha(2s-m-2)\,\dot\alpha}^{(+)}$ be the polarization of the emitting positive helicity higher-spin field. Since $\eps^{(+)}_{\alpha(2s-m-2)\,\dot\alpha}$ transforms as $\delta \eps^{(+)}_{\alpha(2s-m-2)\,\dot\alpha}=k_{\alpha\dot\alpha}\xi_{\alpha(2s-m-3)}$ (after gauge fixing), we can cancel out the propagator in \eqref{sceIstep1} by contracting $k_{\alpha\dot\alpha}$ with $p_i^{\alpha\dot\alpha}$. As a result, Poincar\'e invariance of $\cM_n$ when $k^{\alpha\dot\alpha}\rightarrow 0$ imposes \eqref{claim5}.
\qed

\medskip

Once again, we find that the non-minimal couplings $(+,+,+)$ vanish in the soft limit. In fact, we can even argue why non-minimal couplings will not contribute from the point of view of scattering amplitudes. Suppose we start with an all-plus vertex, then the only vertex it can be glued with is either $\cV_3^{(-,+,+)}$ or $\cV_3^{(-,-,+)}$. As a consequence, we have either $\cM_4(+,+,+,+)$ or $\cM_4(-,+,+,+)$ as 4-pt amplitudes. However, these amplitudes can be shown to vanish on-shell.

\medskip

$\diamond$ \underline{\textbf{Case II:} Implication from soft limit of $\bar{\cF}^{(+)}_{s,m,i}$.} 

Proceed similarly with Case I, we write the soft factor $\bar{\cF}^{(+)}_{s,m,i}$ as:
\be\label{sceIIstep1}
\begin{split}
   \bar{\cF}^{(+)}_{s,m,i}=\tg_{s,m,i}^{-++}\, \frac{[i\,\kappa]^{m+1}\langle n\,i\rangle^{2s-m-1}}{k\cdot p_i\,\langle n\,\kappa\rangle^{2s-m-1}}=\frac{\tg_{s,m,i}^{-++}}{k\cdot p_i} \eps^{(+)}_{\alpha(2s-m-1)\,\dot\alpha}\tilde\rho_i^{\dot\alpha(m+1)}\rho_i^{\alpha(2s-m-1)}\tilde\kappa_{\dot\alpha(m)}\,.
    \end{split}
\ee

\begin{proposition} Gauge invariance of an $n$-point scattering amplitude $\cM_n$ when there is a soft emitting particle with positive helicity field $+s$ imposes:
\begin{align}\label{claim6}
    \sum_i \tg_{s,m,i}^{-++}\,\rho_i^{\alpha(2s-m-2)}\tilde\rho_i^{\dot\alpha(m)}\tilde\kappa_{\dot\alpha(m)}=0\,.
\end{align}
\end{proposition}

\proof Similar to Proposition \ref{prop2}.
\qed

\medskip
\begin{itemize}
    \item[-] When $m=0$, \eqref{claim6} reduces to
    \begin{align}
        \sum_i\tg_{s,0,i}^{-++}\rho_i^{\alpha(2s-2)}=0\,.
    \end{align}
    The solution of the above is $s=1$. As a consequence, we once again obtain charge conservation. Furthermore, this result fits well with the observation in \cite{Adamo:2022lah}. Namely, the choice of $\zeta_{\alpha(2s-1)}$ is not pure gauge unless positive helicity fields have spin-1 when we consider one-derivative interacting theory. This is due to the fact that the difference between two positive helicity higher-spin fields with the same momentum but different $\zeta_{\alpha(2s-1)}$ is not equivalent to a gauge transformation (cf., \eqref{FreeGT}). 
    \item[-] When $m=1$, we have 
    \begin{align}
        \sum_i\tg_{s,1,i}^{-++}\rho_i^{2s-3}\tilde\rho_i^{\dot\alpha}\tilde\kappa_{\dot\alpha}=0\,.
    \end{align}
    Treating $\tilde\kappa$ as constant spinor, we can obtain the equivalent principle by setting $s=2$:
    \begin{align}
       \tilde\kappa_{\dot\alpha} \sum_i\tg_{2,1,i}^{-++}\rho_i^{\alpha}\tilde\rho_i^{\dot\alpha}=0\qquad \Rightarrow \qquad \tg_{2,1,i}^{-++}=const\,.
    \end{align}
\end{itemize}
As alluded to above, the equivalence principle as well as other constraints that come with the higher power of the soft momentum are hidden in the IR.

%%%%%%%%%%%%%%%%%%%%%%%%%%%%%%%%%%%%%%%%%
\section{Discussion}\label{sec:4}

As demonstrated, Weinberg's soft theorem can be avoided with almost no effort when we are willing to abandon parity invariance by working with the chiral representation. In addition, Weinberg's arguments are more tightly related to the number of derivatives in the interactions rather than spins. What surprised us was that all constraints from gauge invariance emerged as we went deeper into the IR, i.e. they are accompanied by higher power of the soft momentum $k^{\alpha\dot\alpha}$ (cf., \eqref{mainresult}). For this reason, (quasi-)chiral higher-spin theories with non-trivial scattering amplitudes can exist regardless the number of derivatives in the interactions, see examples in  \cite{Sharapov:2022faa,Sharapov:2022awp,Bu:2022iak,Tran:2022tft,Adamo:2022lah}. As a result, we should be able to deform away from the chiral sector of chiral HSGRA (and its supersymmetrization \cite{Metsaev:2019aig,Metsaev:2019dqt,Tsulaia:2022csz} thereof) to obtain a quasi-chiral HSGRA with higher-derivative interactions.\footnote{This is perhaps the furthest we can go before a world-sheet description is needed to deal with non-local higher-spin interactions.} 

\medskip

It is worth noting that the Fronsdal representation as well as its dual formulation \cite{Bekaert:2002uh} do not admit a local deformation that can lead to two-derivative (gravitational) interactions. Therefore, at this moment, the chiral representation is the only representation that allows us to construct local theories of higher spin.\footnote{Note that even though higher-spin fields in the chiral representation are complex, they can be used to construct unitary and parity-invariant theory such as GR \cite{Krasnov:2020lku}.} Furthermore, compared with the results of \cite{Benincasa:2007xk,Benincasa:2011kn,Benincasa:2011pg}, we note that we use different assumptions. In particular, we do not assume that the vertices must be parity-invariant, which allows us to construct non-trivial higher $S$-spin matrices.

\medskip

Since our work is heavily based on the chiral representation, it should naturally admit a twistor description in the spirit of \cite{Adamo:2021lrv}. To wit, it may offer a new perspective to flat holography, and more specifically to higher-spin celestial holography~\cite{Ponomarev:2022ryp,Ponomarev:2022qkx}. In \cite{Ponomarev:2017nrr}, a higher-spin generalization of colour-kinematics duality \cite{Monteiro:2011pc} has been studied in the light-cone gauge. It would be interesting to understand the result of \cite{Ponomarev:2017nrr} from a twistor point of view.

\medskip

While we only considered massless higher-spin vertices/amplitudes in this work, it should be possible (and also makes sense) to explore the soft emission of massless particles when there are couplings between massless and massive fields, see e.g. \cite{Conde:2016izb,Metsaev:2022yvb} for a complete classification of 4-dimensional massive/massless higher-spin vertices in the light-cone gauge. This will help us to see if there are other sets of relations which constrain massive-massless higher-spin interactions.

\medskip

%%%%%%%%%%%%%%%%%%%%%%%%%%%%%%%%%%%%%%%%
\acknowledgments
The author appreciates enlightening discussions with Tim Adamo, Chrysoula Markou, Ricardo Monteiro, Zhenya Skvortsov and Harold Steinacker. He also thanks Tim Adamo and Zhenya Skvortsov for useful correspondence. The hospitality of the University of Vienna where this project was initiated is gratefully acknowledged. This research was partially completed at the workshop ``Higher Spin Gravity and its Applications" supported by the Asia Pacific Center for Theoretical Physics. The author is partially supported by the Fonds de la Recherche Scientifique under Grants No. F.4503.20 (HighSpinSymm), Grant No. F.4544.21 (HigherSpinGraWave), and the funding from the European Research Council (ERC) under Grant No. 101002551.
%%%%%%%%%%%%%%%%%%%%%%%%%%%%%%%%%%%%

%%%%%%%%%%%%%%%%%%%%%%%%%%%%%%%%%%%%%%%%

\setstretch{0.8}
\footnotesize
\bibliography{twistor}

\providecommand{\href}[2]{#2}\begingroup\raggedright\begin{thebibliography}{100}

\bibitem{Adamo:2022lah}
T.~Adamo and T.~Tran, {\it {Higher-spin Yang-Mills, amplitudes and
  self-duality}},  \href{http://arxiv.org/abs/2210.07130}{{\tt
  arXiv:2210.07130}}.

\bibitem{Weinberg:1964ew}
S.~Weinberg, {\it {Photons and Gravitons in $S$-Matrix Theory: Derivation of
  Charge Conservation and Equality of Gravitational and Inertial Mass}},  {\em
  Phys. Rev.} {\bf 135} (1964) B1049--B1056.

\bibitem{Coleman:1967ad}
S.~R. Coleman and J.~Mandula, {\it {All Possible Symmetries of the S Matrix}},
  {\em Phys. Rev.} {\bf 159} (1967) 1251--1256.

\bibitem{Maldacena:2011jn}
J.~Maldacena and A.~Zhiboedov, {\it {Constraining Conformal Field Theories with
  A Higher Spin Symmetry}},  {\em J. Phys. A} {\bf 46} (2013) 214011,
  [\href{http://arxiv.org/abs/1112.1016}{{\tt arXiv:1112.1016}}].

\bibitem{Boulanger:2013zza}
N.~Boulanger, D.~Ponomarev, E.~D. Skvortsov, and M.~Taronna, {\it {On the
  uniqueness of higher-spin symmetries in AdS and CFT}},  {\em Int. J. Mod.
  Phys. A} {\bf 28} (2013) 1350162, [\href{http://arxiv.org/abs/1305.5180}{{\tt
  arXiv:1305.5180}}].

\bibitem{Sleight:2017pcz}
C.~Sleight and M.~Taronna, {\it {Higher-Spin Gauge Theories and Bulk
  Locality}},  {\em Phys. Rev. Lett.} {\bf 121} (2018), no.~17 171604,
  [\href{http://arxiv.org/abs/1704.07859}{{\tt arXiv:1704.07859}}].

\bibitem{Weinberg:1995mt}
S.~Weinberg, {\em {The Quantum theory of fields. Vol. 1: Foundations}}.
\newblock Cambridge University Press, 6, 2005.

\bibitem{Bekaert:2010hw}
X.~Bekaert, N.~Boulanger, and P.~Sundell, {\it {How higher-spin gravity
  surpasses the spin two barrier: no-go theorems versus yes-go examples}},
  {\em Rev. Mod. Phys.} {\bf 84} (2012) 987--1009,
  [\href{http://arxiv.org/abs/1007.0435}{{\tt arXiv:1007.0435}}].

\bibitem{Schwartz:2014sze}
M.~D. Schwartz, {\em {Quantum Field Theory and the Standard Model}}.
\newblock Cambridge University Press, 3, 2014.

\bibitem{Strominger:2017zoo}
A.~Strominger, {\it {Lectures on the Infrared Structure of Gravity and Gauge
  Theory}},  \href{http://arxiv.org/abs/1703.05448}{{\tt arXiv:1703.05448}}.

\bibitem{McLoughlin:2022ljp}
T.~McLoughlin, A.~Puhm, and A.-M. Raclariu, {\it {The SAGEX Review on
  Scattering Amplitudes, Chapter 11: Soft Theorems and Celestial Amplitudes}},
  \href{http://arxiv.org/abs/2203.13022}{{\tt arXiv:2203.13022}}.

\bibitem{Bekaert:2022poo}
X.~Bekaert, N.~Boulanger, A.~Campoleoni, M.~Chiodaroli, D.~Francia,
  M.~Grigoriev, E.~Sezgin, and E.~Skvortsov, {\it {Snowmass White Paper: Higher
  Spin Gravity and Higher Spin Symmetry}},
  \href{http://arxiv.org/abs/2205.01567}{{\tt arXiv:2205.01567}}.

\bibitem{Ponomarev:2022vjb}
D.~Ponomarev, {\it {Basic introduction to higher-spin theories}},
  \href{http://arxiv.org/abs/2206.15385}{{\tt arXiv:2206.15385}}.

\bibitem{deMelloKoch:2010wdf}
R.~de~Mello~Koch, A.~Jevicki, K.~Jin, and J.~P. Rodrigues, {\it {$AdS_4/CFT_3$
  Construction from Collective Fields}},  {\em Phys. Rev. D} {\bf 83} (2011)
  025006, [\href{http://arxiv.org/abs/1008.0633}{{\tt arXiv:1008.0633}}].

\bibitem{Bekaert:2015tva}
X.~Bekaert, J.~Erdmenger, D.~Ponomarev, and C.~Sleight, {\it {Quartic AdS
  Interactions in Higher-Spin Gravity from Conformal Field Theory}},  {\em
  JHEP} {\bf 11} (2015) 149, [\href{http://arxiv.org/abs/1508.04292}{{\tt
  arXiv:1508.04292}}].

\bibitem{Boulanger:2015ova}
N.~Boulanger, P.~Kessel, E.~D. Skvortsov, and M.~Taronna, {\it {Higher spin
  interactions in four-dimensions: Vasiliev versus Fronsdal}},  {\em J. Phys.
  A} {\bf 49} (2016), no.~9 095402,
  [\href{http://arxiv.org/abs/1508.04139}{{\tt arXiv:1508.04139}}].

\bibitem{Aharony:2022feg}
O.~Aharony, S.~M. Chester, T.~Sheaffer, and E.~Y. Urbach, {\it {Explicit
  holography for vector models at finite $N$, volume and temperature}},
  \href{http://arxiv.org/abs/2208.13607}{{\tt arXiv:2208.13607}}.

\bibitem{Bern:1993qk}
Z.~Bern, G.~Chalmers, L.~J. Dixon, and D.~A. Kosower, {\it {One loop N gluon
  amplitudes with maximal helicity violation via collinear limits}},  {\em
  Phys. Rev. Lett.} {\bf 72} (1994) 2134--2137,
  [\href{http://arxiv.org/abs/hep-ph/9312333}{{\tt hep-ph/9312333}}].

\bibitem{Mahlon:1993si}
G.~Mahlon, {\it {Multi - gluon helicity amplitudes involving a quark loop}},
  {\em Phys. Rev. D} {\bf 49} (1994) 4438--4453,
  [\href{http://arxiv.org/abs/hep-ph/9312276}{{\tt hep-ph/9312276}}].

\bibitem{Bardeen:1995gk}
W.~A. Bardeen, {\it {Selfdual Yang-Mills theory, integrability and multiparton
  amplitudes}},  {\em Prog. Theor. Phys. Suppl.} {\bf 123} (1996) 1--8.

\bibitem{Bern:1996ja}
Z.~Bern, L.~J. Dixon, D.~C. Dunbar, and D.~A. Kosower, {\it {One loop selfdual
  and N=4 superYang-Mills}},  {\em Phys. Lett. B} {\bf 394} (1997) 105--115,
  [\href{http://arxiv.org/abs/hep-th/9611127}{{\tt hep-th/9611127}}].

\bibitem{Bern:1998sv}
Z.~Bern, L.~J. Dixon, M.~Perelstein, and J.~S. Rozowsky, {\it {Multileg one
  loop gravity amplitudes from gauge theory}},  {\em Nucl. Phys. B} {\bf 546}
  (1999) 423--479, [\href{http://arxiv.org/abs/hep-th/9811140}{{\tt
  hep-th/9811140}}].

\bibitem{Krasnov:2016emc}
K.~Krasnov, {\it {Self-Dual Gravity}},  {\em Class. Quant. Grav.} {\bf 34}
  (2017), no.~9 095001, [\href{http://arxiv.org/abs/1610.01457}{{\tt
  arXiv:1610.01457}}].

\bibitem{Bengtsson:1983pd}
A.~K.~H. Bengtsson, I.~Bengtsson, and L.~Brink, {\it {Cubic Interaction Terms
  for Arbitrary Spin}},  {\em Nucl. Phys. B} {\bf 227} (1983) 31--40.

\bibitem{Bengtsson:1986kh}
A.~K.~H. Bengtsson, I.~Bengtsson, and N.~Linden, {\it {Interacting Higher Spin
  Gauge Fields on the Light Front}},  {\em Class. Quant. Grav.} {\bf 4} (1987)
  1333.

\bibitem{Metsaev:2005ar}
R.~R. Metsaev, {\it {Cubic interaction vertices of massive and massless higher
  spin fields}},  {\em Nucl. Phys. B} {\bf 759} (2006) 147--201,
  [\href{http://arxiv.org/abs/hep-th/0512342}{{\tt hep-th/0512342}}].

\bibitem{Barnich:1993vg}
G.~Barnich and M.~Henneaux, {\it {Consistent couplings between fields with a
  gauge freedom and deformations of the master equation}},  {\em Phys. Lett. B}
  {\bf 311} (1993) 123--129, [\href{http://arxiv.org/abs/hep-th/9304057}{{\tt
  hep-th/9304057}}].

\bibitem{Manvelyan:2010jr}
R.~Manvelyan, K.~Mkrtchyan, and W.~Ruhl, {\it {General trilinear interaction
  for arbitrary even higher spin gauge fields}},  {\em Nucl. Phys. B} {\bf 836}
  (2010) 204--221, [\href{http://arxiv.org/abs/1003.2877}{{\tt
  arXiv:1003.2877}}].

\bibitem{Joung:2013nma}
E.~Joung and M.~Taronna, {\it {Cubic-interaction-induced deformations of
  higher-spin symmetries}},  {\em JHEP} {\bf 03} (2014) 103,
  [\href{http://arxiv.org/abs/1311.0242}{{\tt arXiv:1311.0242}}].

\bibitem{Blencowe:1988gj}
M.~P. Blencowe, {\it {A Consistent Interacting Massless Higher Spin Field
  Theory in $D$ = (2+1)}},  {\em Class. Quant. Grav.} {\bf 6} (1989) 443.

\bibitem{Bergshoeff:1989ns}
E.~Bergshoeff, M.~P. Blencowe, and K.~S. Stelle, {\it {Area Preserving
  Diffeomorphisms and Higher Spin Algebra}},  {\em Commun. Math. Phys.} {\bf
  128} (1990) 213.

\bibitem{Pope:1989vj}
C.~N. Pope and P.~K. Townsend, {\it {Conformal Higher Spin in
  (2+1)-dimensions}},  {\em Phys. Lett. B} {\bf 225} (1989) 245--250.

\bibitem{Fradkin:1989xt}
E.~S. Fradkin and V.~Y. Linetsky, {\it {A Superconformal Theory of Massless
  Higher Spin Fields in $D$ = (2+1)}},  {\em Mod. Phys. Lett. A} {\bf 4} (1989)
  731.

\bibitem{Metsaev:1991mt}
R.~R. Metsaev, {\it {Poincare invariant dynamics of massless higher spins:
  Fourth order analysis on mass shell}},  {\em Mod. Phys. Lett. A} {\bf 6}
  (1991) 359--367.

\bibitem{Metsaev:1991nb}
R.~R. Metsaev, {\it {S matrix approach to massless higher spins theory. 2: The
  Case of internal symmetry}},  {\em Mod. Phys. Lett. A} {\bf 6} (1991)
  2411--2421.

\bibitem{Campoleoni:2010zq}
A.~Campoleoni, S.~Fredenhagen, S.~Pfenninger, and S.~Theisen, {\it {Asymptotic
  symmetries of three-dimensional gravity coupled to higher-spin fields}},
  {\em JHEP} {\bf 11} (2010) 007, [\href{http://arxiv.org/abs/1008.4744}{{\tt
  arXiv:1008.4744}}].

\bibitem{Henneaux:2010xg}
M.~Henneaux and S.-J. Rey, {\it {Nonlinear $W_{infinity}$ as Asymptotic
  Symmetry of Three-Dimensional Higher Spin Anti-de Sitter Gravity}},  {\em
  JHEP} {\bf 12} (2010) 007, [\href{http://arxiv.org/abs/1008.4579}{{\tt
  arXiv:1008.4579}}].

\bibitem{Gaberdiel:2010pz}
M.~R. Gaberdiel and R.~Gopakumar, {\it {An AdS$_{3}$ Dual for Minimal Model
  CFTs}},  {\em Phys. Rev. D} {\bf 83} (2011) 066007,
  [\href{http://arxiv.org/abs/1011.2986}{{\tt arXiv:1011.2986}}].

\bibitem{Gaberdiel:2012uj}
M.~R. Gaberdiel and R.~Gopakumar, {\it {Minimal Model Holography}},  {\em J.
  Phys. A} {\bf 46} (2013) 214002, [\href{http://arxiv.org/abs/1207.6697}{{\tt
  arXiv:1207.6697}}].

\bibitem{Gaberdiel:2014cha}
M.~R. Gaberdiel and R.~Gopakumar, {\it {Higher Spins \& Strings}},  {\em JHEP}
  {\bf 11} (2014) 044, [\href{http://arxiv.org/abs/1406.6103}{{\tt
  arXiv:1406.6103}}].

\bibitem{Grigoriev:2019xmp}
M.~Grigoriev, I.~Lovrekovic, and E.~Skvortsov, {\it {New Conformal Higher Spin
  Gravities in $3d$}},  {\em JHEP} {\bf 01} (2020) 059,
  [\href{http://arxiv.org/abs/1909.13305}{{\tt arXiv:1909.13305}}].

\bibitem{Grigoriev:2020lzu}
M.~Grigoriev, K.~Mkrtchyan, and E.~Skvortsov, {\it {Matter-free higher spin
  gravities in 3D: Partially-massless fields and general structure}},  {\em
  Phys. Rev. D} {\bf 102} (2020), no.~6 066003,
  [\href{http://arxiv.org/abs/2005.05931}{{\tt arXiv:2005.05931}}].

\bibitem{Ponomarev:2016lrm}
D.~Ponomarev and E.~D. Skvortsov, {\it {Light-Front Higher-Spin Theories in
  Flat Space}},  {\em J. Phys. A} {\bf 50} (2017), no.~9 095401,
  [\href{http://arxiv.org/abs/1609.04655}{{\tt arXiv:1609.04655}}].

\bibitem{Metsaev:2018xip}
R.~R. Metsaev, {\it {Light-cone gauge cubic interaction vertices for massless
  fields in AdS(4)}},  {\em Nucl. Phys. B} {\bf 936} (2018) 320--351,
  [\href{http://arxiv.org/abs/1807.07542}{{\tt arXiv:1807.07542}}].

\bibitem{Metsaev:2019dqt}
R.~R. Metsaev, {\it {Cubic interaction vertices for N=1 arbitrary spin massless
  supermultiplets in flat space}},  {\em JHEP} {\bf 08} (2019) 130,
  [\href{http://arxiv.org/abs/1905.11357}{{\tt arXiv:1905.11357}}].

\bibitem{Metsaev:2019aig}
R.~R. Metsaev, {\it {Cubic interactions for arbitrary spin $ \mathcal{N} $
  -extended massless supermultiplets in 4d flat space}},  {\em JHEP} {\bf 11}
  (2019) 084, [\href{http://arxiv.org/abs/1909.05241}{{\tt arXiv:1909.05241}}].

\bibitem{Krasnov:2021nsq}
K.~Krasnov, E.~Skvortsov, and T.~Tran, {\it {Actions for self-dual Higher Spin
  Gravities}},  {\em JHEP} {\bf 08} (2021) 076,
  [\href{http://arxiv.org/abs/2105.12782}{{\tt arXiv:2105.12782}}].

\bibitem{Tran:2022tft}
T.~Tran, {\it {Toward a twistor action for chiral higher-spin gravity}},
  \href{http://arxiv.org/abs/2209.00925}{{\tt arXiv:2209.00925}}.

\bibitem{Tsulaia:2022csz}
M.~Tsulaia and D.~Weissman, {\it {Supersymmetric Quantum Chiral Higher Spin
  Gravity}},  \href{http://arxiv.org/abs/2209.13907}{{\tt arXiv:2209.13907}}.

\bibitem{Herfray:2022prf}
Y.~Herfray, K.~Krasnov, and E.~Skvortsov, {\it {Higher-Spin Self-Dual
  Yang-Mills and Gravity from the twistor space}},
  \href{http://arxiv.org/abs/2210.06209}{{\tt arXiv:2210.06209}}.

\bibitem{Segal:2002gd}
A.~Y. Segal, {\it {Conformal higher spin theory}},  {\em Nucl. Phys. B} {\bf
  664} (2003) 59--130, [\href{http://arxiv.org/abs/hep-th/0207212}{{\tt
  hep-th/0207212}}].

\bibitem{Tseytlin:2002gz}
A.~A. Tseytlin, {\it {On limits of superstring in AdS(5) x S**5}},  {\em Theor.
  Math. Phys.} {\bf 133} (2002) 1376--1389,
  [\href{http://arxiv.org/abs/hep-th/0201112}{{\tt hep-th/0201112}}].

\bibitem{Bekaert:2010ky}
X.~Bekaert, E.~Joung, and J.~Mourad, {\it {Effective action in a higher-spin
  background}},  {\em JHEP} {\bf 02} (2011) 048,
  [\href{http://arxiv.org/abs/1012.2103}{{\tt arXiv:1012.2103}}].

\bibitem{Sperling:2017dts}
M.~Sperling and H.~C. Steinacker, {\it {Covariant 4-dimensional fuzzy spheres,
  matrix models and higher spin}},  {\em J. Phys. A} {\bf 50} (2017), no.~37
  375202, [\href{http://arxiv.org/abs/1704.02863}{{\tt arXiv:1704.02863}}].

\bibitem{Sperling:2018xrm}
M.~Sperling and H.~C. Steinacker, {\it {The fuzzy 4-hyperboloid $H^4_n$ and
  higher-spin in Yang\textendash{}Mills matrix models}},  {\em Nucl. Phys. B}
  {\bf 941} (2019) 680--743, [\href{http://arxiv.org/abs/1806.05907}{{\tt
  arXiv:1806.05907}}].

\bibitem{Steinacker:2022jjv}
H.~Steinacker and T.~Tran, {\it {A Twistorial Description of the IKKT-Matrix
  Model}},  \href{http://arxiv.org/abs/2203.05436}{{\tt arXiv:2203.05436}}.

\bibitem{Mason:2009afn}
L.~J. Mason and D.~Skinner, {\it {Gravity, Twistors and the MHV Formalism}},
  {\em Commun. Math. Phys.} {\bf 294} (2010) 827--862,
  [\href{http://arxiv.org/abs/0808.3907}{{\tt arXiv:0808.3907}}].

\bibitem{Adamo:2020yzi}
T.~Adamo, L.~Mason, and A.~Sharma, {\it {Gluon scattering on self-dual
  radiative gauge fields}},  \href{http://arxiv.org/abs/2010.14996}{{\tt
  arXiv:2010.14996}}.

\bibitem{Adamo:2022mev}
T.~Adamo, L.~Mason, and A.~Sharma, {\it {Graviton scattering in self-dual
  radiative space-times}},  \href{http://arxiv.org/abs/2203.02238}{{\tt
  arXiv:2203.02238}}.

\bibitem{Skvortsov:2018uru}
E.~Skvortsov, {\it {Light-Front Bootstrap for Chern-Simons Matter Theories}},
  {\em JHEP} {\bf 06} (2019) 058, [\href{http://arxiv.org/abs/1811.12333}{{\tt
  arXiv:1811.12333}}].

\bibitem{Sharapov:2022awp}
A.~Sharapov and E.~Skvortsov, {\it {Chiral Higher Spin Gravity in (A)dS${}_4$
  and secrets of Chern-Simons Matter Theories}},
  \href{http://arxiv.org/abs/2205.15293}{{\tt arXiv:2205.15293}}.

\bibitem{Jain:2022ajd}
P.~Jain, S.~Jain, B.~Sahoo, K.~S. Dhruva, and A.~Zade, {\it {Mapping Slightly
  Broken Higher Spin (SBHS) theory correlators to Free theory correlators: A
  momentum space bootstrap using SBHS symmetry}},
  \href{http://arxiv.org/abs/2207.05101}{{\tt arXiv:2207.05101}}.

\bibitem{Penrose:1967wn}
R.~Penrose, {\it {Twistor algebra}},  {\em J. Math. Phys.} {\bf 8} (1967) 345.

\bibitem{Sezgin:2017jgm}
E.~Sezgin, E.~D. Skvortsov, and Y.~Zhu, {\it {Chern-Simons Matter Theories and
  Higher Spin Gravity}},  {\em JHEP} {\bf 07} (2017) 133,
  [\href{http://arxiv.org/abs/1705.03197}{{\tt arXiv:1705.03197}}].

\bibitem{Skvortsov:2022wzo}
E.~Skvortsov and Y.~Yin, {\it {On (spinor)-helicity and bosonization in
  $AdS_4/CFT_3$}},  \href{http://arxiv.org/abs/2207.06976}{{\tt
  arXiv:2207.06976}}.

\bibitem{Skvortsov:2015lja}
E.~D. Skvortsov and M.~Taronna, {\it {On Locality, Holography and Unfolding}},
  {\em JHEP} {\bf 11} (2015) 044, [\href{http://arxiv.org/abs/1508.04764}{{\tt
  arXiv:1508.04764}}].

\bibitem{Fronsdal:1978rb}
C.~Fronsdal, {\it {Massless Fields with Integer Spin}},  {\em Phys. Rev. D}
  {\bf 18} (1978) 3624.

\bibitem{Sharapov:2022faa}
A.~Sharapov, E.~Skvortsov, A.~Sukhanov, and R.~Van~Dongen, {\it {Minimal model
  of Chiral Higher Spin Gravity}},  \href{http://arxiv.org/abs/2205.07794}{{\tt
  arXiv:2205.07794}}.

\bibitem{Conde:2016izb}
E.~Conde, E.~Joung, and K.~Mkrtchyan, {\it {Spinor-Helicity Three-Point
  Amplitudes from Local Cubic Interactions}},  {\em JHEP} {\bf 08} (2016) 040,
  [\href{http://arxiv.org/abs/1605.07402}{{\tt arXiv:1605.07402}}].

\bibitem{Roiban:2017iqg}
R.~Roiban and A.~A. Tseytlin, {\it {On four-point interactions in massless
  higher spin theory in flat space}},  {\em JHEP} {\bf 04} (2017) 139,
  [\href{http://arxiv.org/abs/1701.05773}{{\tt arXiv:1701.05773}}].

\bibitem{Lysov:2022nsv}
V.~Lysov and Y.~Neiman, {\it {Bulk locality and gauge invariance for
  boundary-bilocal cubic correlators in higher-spin gravity}},
  \href{http://arxiv.org/abs/2209.00854}{{\tt arXiv:2209.00854}}.

\bibitem{Neiman:2022enh}
Y.~Neiman, {\it {New diagrammatic framework for higher-spin gravity}},
  \href{http://arxiv.org/abs/2209.02185}{{\tt arXiv:2209.02185}}.

\bibitem{Tran:2021ukl}
T.~Tran, {\it {Twistor constructions for higher-spin extensions of (self-dual)
  Yang-Mills}},  {\em JHEP} {\bf 11} (2021) 117,
  [\href{http://arxiv.org/abs/2107.04500}{{\tt arXiv:2107.04500}}].

\bibitem{Krasnov:2020lku}
K.~Krasnov, {\em {Formulations of General Relativity}}.
\newblock Cambridge Monographs on Mathematical Physics. Cambridge University
  Press, 11, 2020.

\bibitem{Cachazo:2014fwa}
F.~Cachazo and A.~Strominger, {\it {Evidence for a New Soft Graviton Theorem}},
   \href{http://arxiv.org/abs/1404.4091}{{\tt arXiv:1404.4091}}.

\bibitem{Bloch:1937pw}
F.~Bloch and A.~Nordsieck, {\it {Note on the Radiation Field of the electron}},
   {\em Phys. Rev.} {\bf 52} (1937) 54--59.

\bibitem{Joung:2015eny}
E.~Joung, S.~Nakach, and A.~A. Tseytlin, {\it {Scalar scattering via conformal
  higher spin exchange}},  {\em JHEP} {\bf 02} (2016) 125,
  [\href{http://arxiv.org/abs/1512.08896}{{\tt arXiv:1512.08896}}].

\bibitem{Beccaria:2016syk}
M.~Beccaria, S.~Nakach, and A.~A. Tseytlin, {\it {On triviality of S-matrix in
  conformal higher spin theory}},  {\em JHEP} {\bf 09} (2016) 034,
  [\href{http://arxiv.org/abs/1607.06379}{{\tt arXiv:1607.06379}}].

\bibitem{Skvortsov:2018jea}
E.~D. Skvortsov, T.~Tran, and M.~Tsulaia, {\it {Quantum Chiral Higher Spin
  Gravity}},  {\em Phys. Rev. Lett.} {\bf 121} (2018), no.~3 031601,
  [\href{http://arxiv.org/abs/1805.00048}{{\tt arXiv:1805.00048}}].

\bibitem{Skvortsov:2020wtf}
E.~Skvortsov, T.~Tran, and M.~Tsulaia, {\it {More on Quantum Chiral Higher Spin
  Gravity}},  {\em Phys. Rev. D} {\bf 101} (2020), no.~10 106001,
  [\href{http://arxiv.org/abs/2002.08487}{{\tt arXiv:2002.08487}}].

\bibitem{Skvortsov:2020gpn}
E.~Skvortsov and T.~Tran, {\it {One-loop Finiteness of Chiral Higher Spin
  Gravity}},  {\em JHEP} {\bf 07} (2020) 021,
  [\href{http://arxiv.org/abs/2004.10797}{{\tt arXiv:2004.10797}}].

\bibitem{Ponomarev:2017nrr}
D.~Ponomarev, {\it {Chiral Higher Spin Theories and Self-Duality}},  {\em JHEP}
  {\bf 12} (2017) 141, [\href{http://arxiv.org/abs/1710.00270}{{\tt
  arXiv:1710.00270}}].

\bibitem{Kaparulin:2012px}
D.~S. Kaparulin, S.~L. Lyakhovich, and A.~A. Sharapov, {\it {Consistent
  interactions and involution}},  {\em JHEP} {\bf 01} (2013) 097,
  [\href{http://arxiv.org/abs/1210.6821}{{\tt arXiv:1210.6821}}].

\bibitem{Haehnel:2016mlb}
P.~H\"ahnel and T.~McLoughlin, {\it {Conformal higher spin theory and twistor
  space actions}},  {\em J. Phys. A} {\bf 50} (2017), no.~48 485401,
  [\href{http://arxiv.org/abs/1604.08209}{{\tt arXiv:1604.08209}}].

\bibitem{Adamo:2016ple}
T.~Adamo, P.~H\"ahnel, and T.~McLoughlin, {\it {Conformal higher spin
  scattering amplitudes from twistor space}},  {\em JHEP} {\bf 04} (2017) 021,
  [\href{http://arxiv.org/abs/1611.06200}{{\tt arXiv:1611.06200}}].

\bibitem{Ren:2022sws}
L.~Ren, M.~Spradlin, A.~Yelleshpur~Srikant, and A.~Volovich, {\it {On effective
  field theories with celestial duals}},  {\em JHEP} {\bf 08} (2022) 251,
  [\href{http://arxiv.org/abs/2206.08322}{{\tt arXiv:2206.08322}}].

\bibitem{Monteiro:2022lwm}
R.~Monteiro, {\it {Celestial chiral algebras, colour-kinematics duality and
  integrability}},  \href{http://arxiv.org/abs/2208.11179}{{\tt
  arXiv:2208.11179}}.

\bibitem{Damour:1987fp}
T.~Damour and S.~Deser, {\it {Higher Derivative Interactions of Higher Spin
  Gauge Fields}},  {\em Class. Quant. Grav.} {\bf 4} (1987) L95.

\bibitem{Britto:2005fq}
R.~Britto, F.~Cachazo, B.~Feng, and E.~Witten, {\it {Direct proof of tree-level
  recursion relation in Yang-Mills theory}},  {\em Phys. Rev. Lett.} {\bf 94}
  (2005) 181602, [\href{http://arxiv.org/abs/hep-th/0501052}{{\tt
  hep-th/0501052}}].

\bibitem{Chalmers:1996rq}
G.~Chalmers and W.~Siegel, {\it {The Selfdual sector of QCD amplitudes}},  {\em
  Phys. Rev. D} {\bf 54} (1996) 7628--7633,
  [\href{http://arxiv.org/abs/hep-th/9606061}{{\tt hep-th/9606061}}].

\bibitem{Rodina:2018pcb}
L.~Rodina, {\it {Scattering Amplitudes from Soft Theorems and Infrared
  Behavior}},  {\em Phys. Rev. Lett.} {\bf 122} (2019), no.~7 071601,
  [\href{http://arxiv.org/abs/1807.09738}{{\tt arXiv:1807.09738}}].

\bibitem{Casali:2014xpa}
E.~Casali, {\it {Soft sub-leading divergences in Yang-Mills amplitudes}},  {\em
  JHEP} {\bf 08} (2014) 077, [\href{http://arxiv.org/abs/1404.5551}{{\tt
  arXiv:1404.5551}}].

\bibitem{Bu:2022iak}
W.~Bu, S.~Heuveline, and D.~Skinner, {\it {Moyal deformations, $W_{1+\infty}$
  and celestial holography}},  \href{http://arxiv.org/abs/2208.13750}{{\tt
  arXiv:2208.13750}}.

\bibitem{Bekaert:2002uh}
X.~Bekaert, N.~Boulanger, and M.~Henneaux, {\it {Consistent deformations of
  dual formulations of linearized gravity: A No go result}},  {\em Phys. Rev.
  D} {\bf 67} (2003) 044010, [\href{http://arxiv.org/abs/hep-th/0210278}{{\tt
  hep-th/0210278}}].

\bibitem{Benincasa:2007xk}
P.~Benincasa and F.~Cachazo, {\it {Consistency Conditions on the S-Matrix of
  Massless Particles}},  \href{http://arxiv.org/abs/0705.4305}{{\tt
  arXiv:0705.4305}}.

\bibitem{Benincasa:2011kn}
P.~Benincasa and E.~Conde, {\it {On the Tree-Level Structure of Scattering
  Amplitudes of Massless Particles}},  {\em JHEP} {\bf 11} (2011) 074,
  [\href{http://arxiv.org/abs/1106.0166}{{\tt arXiv:1106.0166}}].

\bibitem{Benincasa:2011pg}
P.~Benincasa and E.~Conde, {\it {Exploring the S-Matrix of Massless
  Particles}},  {\em Phys. Rev. D} {\bf 86} (2012) 025007,
  [\href{http://arxiv.org/abs/1108.3078}{{\tt arXiv:1108.3078}}].

\bibitem{Adamo:2021lrv}
T.~Adamo, L.~Mason, and A.~Sharma, {\it {Celestial $w_{1+\infty}$ Symmetries
  from Twistor Space}},  {\em SIGMA} {\bf 18} (2022) 016,
  [\href{http://arxiv.org/abs/2110.06066}{{\tt arXiv:2110.06066}}].

\bibitem{Ponomarev:2022ryp}
D.~Ponomarev, {\it {Towards higher-spin holography in flat space}},
  \href{http://arxiv.org/abs/2210.04035}{{\tt arXiv:2210.04035}}.

\bibitem{Ponomarev:2022qkx}
D.~Ponomarev, {\it {Chiral higher-spin holography in flat space: the
  Flato-Fronsdal theorem and lower-point functions}},
  \href{http://arxiv.org/abs/2210.04036}{{\tt arXiv:2210.04036}}.

\bibitem{Monteiro:2011pc}
R.~Monteiro and D.~O'Connell, {\it {The Kinematic Algebra From the Self-Dual
  Sector}},  {\em JHEP} {\bf 07} (2011) 007,
  [\href{http://arxiv.org/abs/1105.2565}{{\tt arXiv:1105.2565}}].

\bibitem{Metsaev:2022yvb}
R.~R. Metsaev, {\it {Interacting massive and massless arbitrary spin fields in
  4d flat space}},  {\em Nucl. Phys. B} {\bf 984} (2022) 115978,
  [\href{http://arxiv.org/abs/2206.13268}{{\tt arXiv:2206.13268}}].

\end{thebibliography}\endgroup
\bibliographystyle{JHEP}

\end{document}